\documentclass[12pt, draftclsnofoot, onecolumn]{IEEEtran}
\usepackage[export]{adjustbox}
\usepackage{stfloats}
\usepackage{float}
\usepackage{amsthm}
\usepackage{graphicx}
\usepackage{enumerate}
\usepackage{amsmath}
\usepackage{amssymb}
\usepackage{amsfonts}
\usepackage{filecontents}
\usepackage{xcolor}
\usepackage{bm}
\usepackage{cite}
\usepackage{mathtools}
\usepackage[ruled]{algorithm2e}
\theoremstyle{definition}
\newtheorem{problem}{Problem}
\newtheorem{definition}{Definition}
\newtheorem{lemma}{Lemma}
\newtheorem{theorem}{Theorem}
\newtheorem{remark}{Remark}

\ifCLASSOPTIONcompsoc
\usepackage[caption=false,font=normalsize,labelfon
t=sf,textfont=sf,subrefformat=parens,labelformat=parens]{subfig}
\else
\usepackage[caption=false,font=footnotesize,subrefformat=parens,labelformat=parens]{subfig}
\fi

\title{Computing and Communicating Functions in Disorganized Wireless Networks}
\author{Fangzhou Wu, Li Chen, Nan Zhao, \IEEEmembership{Senior Member, IEEE,} Yunfei Chen, \IEEEmembership{Senior Member, IEEE,} F. Richard Yu, \IEEEmembership{Fellow, IEEE,} and Guo Wei
	\thanks{F. Wu, L. Chen and G. Wei are with Department of Electronic Engineering and Information Science, University of Science and Technology of China, Hefei, Anhui 230027. (e-mail: fangzhouwu@outlook.com, \{chenli87, wei\}@ustc.edu.cn).}
	\thanks{N. Zhao is with the School of Info. and Commun. Eng., Dalian University of Technology, Dalian 116024, China, and also with National Mobile Communications Research Laboratory, Southeast University, Nanjing 210096, China. (e-mail:zhaonan@dlut.edu.cn).}
	\thanks{Y. Chen is with the School of Engineering, University of Warwick, Coventry CV4 7AL, U.K. (e-mail: Yunfei.Chen@warwick.ac.uk).}
	\thanks{F.R. Yu is with the Department of Systems and Computer Engineering, Carleton University, Ottawa, ON, K1S 5B6, Canada. (email: richard.yu@carleton.ca).}
}

\begin{document}
	\maketitle
	\begin{abstract}
		For future wireless networks, enormous numbers of interconnections are required, creating a disorganized topology and leading to a great challenge in data aggregation. Instead of collecting data individually, a more efficient technique, computation over multi-access channels (CoMAC), has emerged to compute functions by exploiting the signal-superposition property of wireless channels. However, the implementation of CoMAC in disorganized networks with multiple relays (hops) is still an open problem. In this paper, we combine CoMAC and orthogonal communication in the disorganized network to attain the computation of functions at the fusion center. First, to make the disorganized network more tractable, we reorganize the disorganized network into a hierarchical network with multiple layers that consists of subgroups and groups. In the hierarchical network, we propose multi-layer function computation where CoMAC is applied to each subgroup and orthogonal communication is  adopted within each group. By computing and communicating subgroup and group functions over layers, the desired functions are reconstructed at the fusion center. The general computation rate is derived and the performance is further improved through time allocation and power control. The closed-form solutions to optimization are obtained, which suggest that existing CoMAC and orthogonal communication schemes can be generalized.	
		
	\end{abstract}
	
	
	\section{Introduction}
	5G and Internet of Things lead to a revolution in wireless networks \cite{fettweis20145g,al2015internet}. With such enormous numbers of nodes, the typologies of networks become complex and disorganized. To aggregate a large amount of data wirelessly in disorganized networks, the conventional multi-access schemes cannot be applied since this would result in excessive network latency with limited radio resources. Thus, how to aggregate data efficiently from distributed nodes is of great importance.
	
	Data aggregation from distributed nodes is first constructed as information-theoretic formulations \cite{orlitsky1995coding,xie2004network,giridhar2005computing,xie2007multisource,moscibroda2007worst,huang2015upper}. For example, a source coding problem involving communicating a function of two variables in a simple two-node network with side information at the receiver, has been solved in  \cite{orlitsky1995coding}. Considering the multi-source network,  \cite{xie2004network} described it as one of communicating possibly correlated sources over a multi-terminal wireless network. Further, given different typology for the disorganized network including relays,  \cite{giridhar2005computing,xie2007multisource,moscibroda2007worst,huang2015upper} provided the corresponding bound of the capacity. Although these interesting bounds were presented in terms of different wireless networks in the ideal case, the study is still limited in the practical case of data aggregation considering channel fading, noise, and resource allocation.
	
	Considering practical wireless networks, some data aggregation approaches introduced different protocols to improve the efficiency of data aggregation,  such as mobile data collectors  \cite{luo2010joint}, topology control  \cite{han2012minimum}, and sleep schedule  \cite{han2013duty}. In these data aggregation schemes, each node at the edge has to transmit its data to the fusion center through several relays (hops). For the disorganized network, the cost of direct data forwarding will be high and the routing path becomes complex due to massive numbers of sources and relays. To further improve the efficiency of data aggregation, compressive sensing was used in the disorganized network to process the data before transmission. In  \cite{roughan2012spatio,kong2013data}, compressive sensing was only applied to remove temporal redundancy before data transmission starts, which did not leverage the flexibly of relays in the routing path. A more precise scheme was proposed by  \cite{xu2015hierarchical}, which aimed to remove data redundancy existing in the routing path and was based on a multilevel hierarchical clustering architecture and hybrid compressive sensing. Unfortunately, even though compressive sensing provides an attracting way to improve the efficiency, the number of measurements at each relay still becomes large in the network with enormous numbers of nodes. This implies that only limited improvement can be obtained by orthogonal communication.
	
	Recently, computation over multi-access channels (CoMAC) has emerged as a promising solution that merges computation and communication by exploiting the signal-superposition property of wireless channels. It collects a relevant function of the node measurements via concurrent node transmissions instead of individual data \cite{goldenbaum2015nomographic,abari2016over,goldenbaum2014channel,nazer2007computation,appuswamy2014computing,nazer2011compute,jeon2014computation,wu2019computation,kortke2014analog,chen2018over2,erez2005lattices,abari2015airshare,jeon2016opportunistic}. These functions computed by CoMAC belong to a class of nomographic functions such as averaging and geometric mean, and are widely used in data aggregation \cite{goldenbaum2015nomographic}. As a straightforward use of CoMAC, nodes in wireless sensor networks can transmit their readings over the air simultaneously to compute a function value of the sensor readings (e.g., arithmetic mean, polynomial or the number of active nodes) instead of requiring individual readings. 
	
	CoMAC was first studied in  \cite{abari2016over,goldenbaum2014channel}, where pre-processing at each node and post-processing at the fusion center were used to compute functions against the fading channel. The designs of pre-processing and post-processing used to compute linear and non-linear functions have been proposed in  \cite{abari2016over}, and the effect of channel estimation error was characterized in  \cite{goldenbaum2014channel}. For robustness to noise, CoMAC was further proposed using joint source-channel coding in  \cite{nazer2007computation,appuswamy2014computing,erez2005lattices,nazer2011compute,jeon2014computation,wu2019computation,jeon2016opportunistic,chen2018over2,kortke2014analog,abari2015airshare} to improve the equivalent signal-to-noise ratio (SNR). The potential of linear source coding was discussed in  \cite{nazer2007computation}, and its application for CoMAC was presented in  \cite{appuswamy2014computing}. Compared with linear source coding, nested lattice coding could approach the performance of a standard random coding  \cite{erez2005lattices}. The lattice-based CoMAC was extended to a general framework in  \cite{nazer2011compute} for networks with linear channels and additive white Gaussian noise. In  \cite{jeon2014computation}, the authors derived the corresponding achievable computation rate considering channel fading. The scheme based on function division was given in  \cite{jeon2016opportunistic,wu2019computation} through theoretical analysis. For the implementation of CoMAC, frequency synchronization has been solved by an attractive solution, called ``AirShare'', which was developed in  \cite{abari2015airshare} for synchronizing nodes by broadcasting a reference-clock signal. To cope with phase offset, a design has been proposed to estimate phase offset and to equalize the corresponding error in  \cite{chen2018over2}. To verify the feasibility of CoMAC in practice, software-defined radio was built in  \cite{kortke2014analog}, and the authors in  \cite{chen2018over2} implemented a cooperative wide-band spectrum sensing system.
	
	However, CoMAC has only been investigated in relay-free networks through direct communications and it cannot be used in the disorganized network where the transmission destination of each node is different. In disorganized networks, analyses  were constructed as information-theoretic formulations in the ideal case without fading channel, noise, and resource allocation. Thus, how to compute functions in practical disorganized networks needs to be addressed. Since the topology becomes more general but also disorganized, it analysis would be quite complex compared with that of the relay-free network.
	
	Motivated by the above observations, in this paper, we first recast the disorganized network into a hierarchical network with multiple layers consisting of subgroups and groups for further analysis. Then, in the hierarchical network, we propose multi-layer function computation (ML-FC) by computing and communicating subgroup and group functions over layers and reconstructing the desired function at the fusion center. Theoretical expressions of achievable computation rates are derived based on nested lattice coding. Furthermore, resource allocation is considered to improve the computation rate, and the corresponding closed-form solutions are given in different cases. Our contributions are summarized as follows: 
	
	\begin{itemize}
		\item \emph{Hierarchical networks}.  We reorganize the disorganized network into the hierarchical networks with multiple layers by introducing two components, i.e., groups and subgroups. Referring to disorganized networks, in hierarchical networks, the nodes in the first layer are source nodes, the only one node in the last layer is the fusion center, and the nodes in the rest layers are relay nodes.
		\item \emph{ML-FC}. In the hierarchical network, ML-FC is proposed which combines CoMAC and orthogonal communication. The desired functions at the fusion center are reconstructed by subgroup functions and group functions where each subgroup function is obtained by CoMAC and the group function is obtained by orthogonal communication.
		\item \emph{General computation rate}. The theoretical expression of the computation rate of ML-FC is derived, and it suggests that the subgroup with the worst computation rate plays an important role in the network. Also, it generalizes CoMAC considering the relay-free network.
		\item \emph{Time allocation and power control}. We formulate two optimization problems considering time allocation with fixed power control and adaptive power control, respectively. Both closed-form solutions are derived, which suggests that the performance with adaptive power control is further improved compared with fixed power control.
	\end{itemize}
	\begin{figure}
		\centering
		\includegraphics[width=0.7\linewidth]{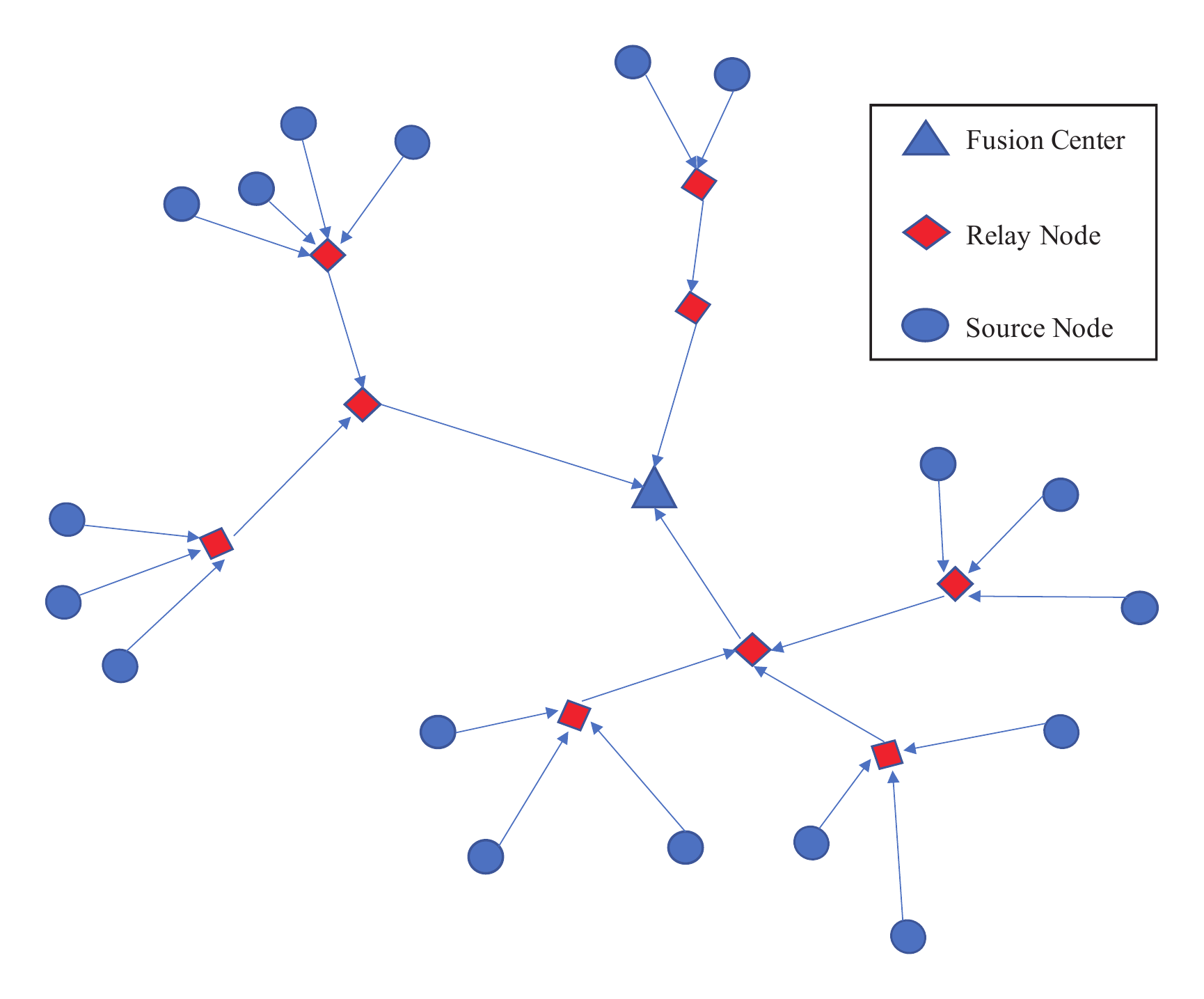}
		\caption{The topology of the disorganized network.}
		\label{fig:randomnetwork}
	\end{figure}
	
	The rest of the paper is organized as follows. Section \ref{System Model??} introduces the considered disorganized network and the classical schemes of data aggregation. The reorganization of the disorganized network and the structure of the hierarchical network are shown in Section \ref{Reorganization of Disorganized Networks}. Based on the hierarchical network, we provide a detailed description of ML-FC in Section \ref{Computation Rate of Multi-layer Networks}, and the computation rate of ML-FC is also derived. Section \ref{Power Control and Time Allocatio} focuses on the analysis of the performance of the proposed ML-FC, which includes power control and time allocation. Simulation results and the corresponding discussions are presented in Section \ref{Simulation Results and Discussions}, and conclusions are given in Section \ref{Conclusion}.
	
	$ \bm{\mathrm{Notations:}} $ Throughout this paper, we define $ \mathsf{C}(x)=\log(1+x) $ and $ \mathsf{C}^+(x)=\max\left\lbrace \frac{1}{2}\log(x),0\right\rbrace$. Let $  [1:n] $ denote a set $ \left\lbrace1,2,\cdots,n\right\rbrace$. For a set $ \mathcal{A} $, $ \left|\mathcal{A}\right|  $  denotes the cardinality of $ \mathcal{A} $. Let the entropy of a random variable $ A $ be $ H(A) $ and the expectation of it be $ \mathsf{E}\left[A \right]  $. A set $ \left\lbrace x_1, x_2, \cdots, x_N \right\rbrace  $ is written as $ \left\lbrace x_i \right\rbrace_{i\in[1:N]}  $ or $ \left\lbrace x_i \right\rbrace_{i=1}^{N} $ for short.    
	
	\section{System Model}\label{System Model??}
	In this section, we first present the topology of the disorganized network. Then, classical aggregation schemes are introduced. Based on the features of the disorganized network, we raise some open problems regarding function computation in the disorganized network.
	\subsection{Disorganized Networks}

	
	In practical terms, the topology of a wireless network is arbitrary. With different deployment environment, the network would be disorganized. Thus, we consider the disorganized network in the general case\footnote{The topology of the wireless network can be arbitrary, but it must be known.}, which consists of source nodes, relay nodes, and one fusion center as destination. In the disorganized network, the fusion center wishes to compute the desired function concerning all the source nodes. With a given topology, the network is demonstrated as Fig.~\ref{fig:randomnetwork}. We assume the number of the source nodes is $ K_1 $ and define a set $ \mathcal{K}_{1} $ including the indexes of all the source nodes. The $ i $-th source node $ {\rm N}_{1,i} $ draws data from the corresponding random source $ S_i $ for $ T_d $ times and then provides a length-$ T_d $ data vector as $ \bm{\mathrm{s}}_{1,i}=[ s_{1,i}[1],\cdots, s_{1,i}[j],\cdots s_{1,i}[T_d]]$.
	
	Let $ \bm{\mathrm{b_v}}=\left[ S_1, S_2, \cdots, S_K\right]  $ be the random source vector associated with a joint probability mass function $ p_{\bm{\mathrm{b_v}}}(\cdot) $. The desired function determined by the random source vector $ \bm{\mathrm{b_v}} $ is expressed as $ f(\bm{\mathrm{b_v}}) $, and its definition is given as follows.
	
	\begin{definition}[Desired Function]
		For all $ j\in[1:T_d] $, the function with independent variables $ \lbrace s_{1,1}[j],s_{1,2}[j],\cdots,s_{1,K_1}[j]\rbrace $ is called the desired function with the form as
		\begin{equation}
		f(s_{1,1}[j],s_{1,2}[j],\cdots,s_{1,K_1}[j])=f(\bm{\mathrm{s}}_1[j]),
		\end{equation}
		where $\bm{\mathrm{s}}_1[j]=[s_{1,1}[j],s_{1,2}[j],\cdots,s_{1,K_1}[j]]$ is independently drawn from $ p_{\bm{\mathrm{b_v}}}(\cdot) $. Every function $ f(\bm{\mathrm{s}}_1[j]) $ is seen as a realization of $  f(\bm{\mathrm{b_v}}) $. Thus, the fusion center computes $ T_d $ desired functions when each source node gets data from each random source for $ T_d $ times. 
	\end{definition}
	
	\begin{remark}[Typical Desired Functions]\label{Typical Functions}
		As studied in  \cite{goldenbaum2014computation,jeon2014computation}, CoMAC is designed to compute different types of desired functions. There are two typical functions that we focus on. The function $ f(\bm{\mathrm{s}}_1[j])$, with values in the set $ \lbrace \sum_{i=1}^{K_1}a_{1,i}s_{1,i}[j], \cdots, \sum_{i=1}^{K_1}a_{L_s,i}s_{1,i}[j] \rbrace  $, is called the arithmetic sum function, where $ a_{l,i}\in\mathbb{R} $ is the weighting factor for the node $ {\rm N}_{1,i} $, and $ L_s $ belongs to $ \mathbb{N} $. The arithmetic sum function is a weighted sum function, which includes the mean function $ f(\bm{\mathrm{s}}_1[j])=\frac{1}{K_1}\sum_{i=1}^{K_1}s_{1,i}[j] $ and the function for the active node only $ f(\bm{\mathrm{s}}_1[j])=\left\lbrace s_{1,1}[j], s_{1,2}[j], \cdots, s_{1,K_1}[j]\right\rbrace $ as special cases. Otherwise, the function $ f(\bm{\mathrm{s}}_1[j])$, with values in the set of $\lbrace \sum_{i=1}^{K_1}\bm{1}_{s_{1,i}[j]=0}, \cdots,  \sum_{i=1}^{K_1}\bm{1}_{s_{1,i}[j]=p}\rbrace  $, is regarded as the type function where $ \bm{1}_{(\cdot)} $ denotes the indicator function and $ p\in\mathbb{N} $. As pointed out in  \cite{giridhar2005computing}, any symmetric function such as mean, variance, maximum, minimum and median can be attained from the type function.
	\end{remark}

	To attain reliable computations against noise, a block code is used, named sequences of nested lattice codes \cite{nazer2011compute}. With the length-$ \bar{n} $ block code, the computation rate is used as performance metrics \cite{jeon2014computation,goldenbaum2015nomographic,goldenbaum2014computation,wu2019computation,nazer2011compute}, of which the definition is given as follows.
	
	\begin{definition}[Computation Rate]\label{Computation rate}
		The computation rate specifies how many function values can be computed per channel use within a predefined accuracy. It can be written as $ R=\lim\limits_{n\rightarrow \infty}\frac{T_d}{n}H(f(\bm{\mathrm{b_v}})) $, where $ T_d $ is the number of function values, $ n $ ($ n\ge\bar{n} $) is the number of channel uses \footnote{If the number of channel uses is equal to the length of the block code, then the computation rate $ R $ is also given as  $ R=\lim\limits_{n\rightarrow \infty}\frac{T_d}{\bar{n}}H(f(\bm{\mathrm{b_v}})) $.} and $ H(f(\bm{\mathrm{b_v}})) $ is the entropy of $ f(\bm{\mathrm{b_v}})  $. Otherwise, $ R $ is achievable only if there is a length-$ \bar{n} $ block code so that the probability $ \Pr\left( \bigcup_{j=1}^{T_d}\left\lbrace \hat{f}(\bm{\mathrm{s}}[j]\neq f(\bm{\mathrm{s}}[j]))\right\rbrace\right)  \rightarrow0 $ as $ \bar{n} $ increases, where $ \hat{f}(\bm{\mathrm{s}}[j]) $ is the estimated function.
	\end{definition}

	\subsection{Aggregation Schemes} \label{Aggregation Schemes}
	
	There exist two classical aggregation schemes, namely CoMAC and orthogonal communication.
	
	\begin{figure}
		\centering
		\includegraphics[width=0.7\linewidth]{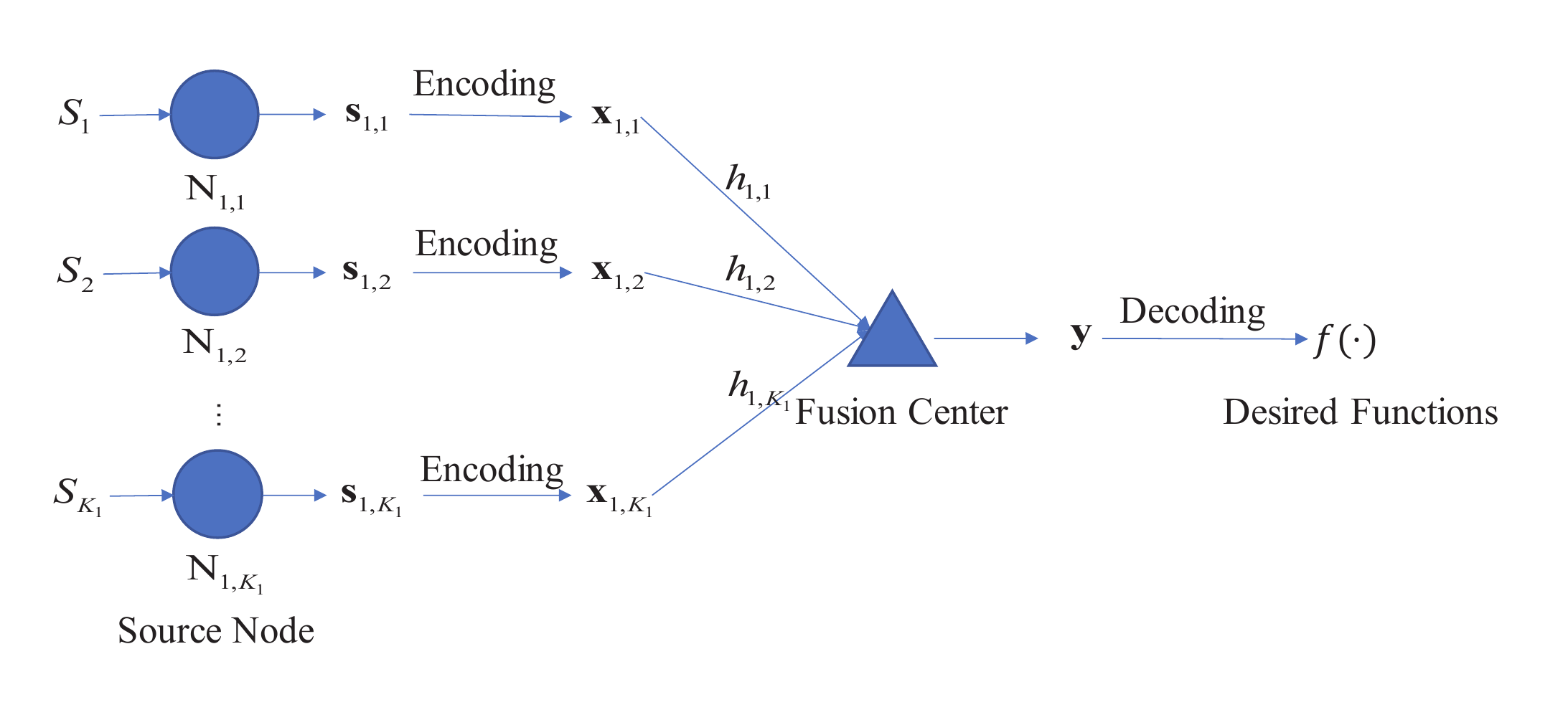}
		\caption{The classical CoMAC for the relay-free network}
		\label{fig:singlelayer}
	\end{figure}
	
	\begin{itemize}
		\item \textbf{\emph{CoMAC.}} CoMAC has been well investigated as the efficient aggregation in the relay-free network (the simplest network), and its classical framework is given in Fig.~\ref{fig:singlelayer}. Different from the disorganized network, in Fig.~\ref{fig:singlelayer}, all the source nodes and the fusion center can be communicated with each other directly.
		
		Let $ \bm{\mathrm{s}}_{1,i} $ represent the data vector of the node $ {\rm N}_{1,i} $ whose length is $ T_d $. Denote $ \bm{\mathrm{x}}_{1,i}=[x_{1,i}[1],x_{1,i}[2],\cdots,x_{1,i}[\bar{n}]] $ as the length-$ \bar{n} $ transmitted vector of the node $ {\rm N}_{1,i} $. The univariate function $ \bm{\mathcal{E}}_{1,i}(\cdot) $ which generates $ \bm{\mathrm{x}}_{1,i}=\bm{\mathcal{E}}_{1,i}(\bm{\mathrm{s}}_{1,i}) $ is an encoding function of $ \mathrm{N}_{1,i} $. This means that $ \bm{\mathrm{s}}_{1,i} $ with length $ T_d $ is mapped to a transmitted vector $ \bm{\mathrm{x}}_{1,i} $ with length $ \bar{n} $ for $ \mathrm{N}_{1,i} $. Then, the received signal for the $ m $-th channel use can be expressed as
		\begin{equation}\label{model of snetwork}
		y[m]=\sum_{i=1}^{{K}_{1}}v_{1,i}[m]h_{1,i}[m]x_{1,i}[m]+w[m],
		\end{equation}
		where $ h_{1,i}[m] $ is the channel from $ \mathrm{N}_{1,i} $ to the fusion center at the $ m $-th channel use, $ x_{1,i}[m] $ is the $ m $-th element of the transmitted vector $ \bm{\mathrm{x}}_{1,i} $, $ v_{1,i}[m]=\frac{|h_{1,i}[m]|}{h_{1,i}[m]}\sqrt{P_{1,i}[m]} $ is the power factor of $ \mathrm{N}_{1,i} $, $ P_{1,i}[m] $ is the transmitted power of $ \mathrm{N}_{1,i} $ and $ w[m] $ is identically and independently distributed (i.i.d.) complex Gaussian random noise following $ \mathcal{CN}(0,1) $.
		
		After $ \bar{n} $ channel uses, the received vector $ \bm{\mathrm{y}} $ is obtained at the fusion center. The decoding function $ \bm{\mathcal{D}}_j(\cdot) $ is used to estimate the $ j $-th desired function $  f(\bm{\mathrm{s}}_1[j])  $, which satisfies $ \hat{f}(\bm{\mathrm{s}}_1[j])=\bm{\mathcal{D}}_j(\bm{\mathrm{y}}) $. This implies that the fusion center obtains $ T_d $ desired functions depending on the received vector with length $ \bar{n} $. Its computation rate  \cite[Theorem 3]{jeon2014computation} is given as
		\begin{equation}\label{RF-CoMAC with FPC}
		R= \mathsf{C}^{+}\left(\dfrac{1}{K_1}+ \mathsf{E}\left[\min_{i\in[1:K_1]}\left|h_{1,i} \right|^2 \right]P  \right) ,
		\end{equation}
		where $ K_1 $ is the number of source nodes in the network, $ P $ is the transmitted power of each node and $ \left|h_{1,i} \right|^2 $ is the channel gain from $ \mathrm{N}_{1,i} $ to the fusion center.
		\item \textbf{\emph{Orthogonal Communication.}} The other solution to aggregating data uses orthogonal resource blocks (e.g., channel uses, code sequences, and sub-carriers) to transmit the individual data to the fusion center. To compute the $j $-th desired function $ f(\bm{\mathrm{s}}_1[j]) $, the fusion center should first obtain the individual data $ \{s_{1,i}[j]\}_{i=1}^{K_1} $ during $ K_1 $ channel uses. Then, the corresponding desired function $ f(\bm{\mathrm{s}}_1[j]) $ is calculated. It is also known as the time-sharing technique, which achieves a computation rate of
		\begin{equation}\label{the time-sharing technique}
		R=\dfrac{1}{K_1} \mathsf{C}\left(\mathsf{E}\left[\left|h \right|^2\right]P  \right),
		\end{equation}
		where  $ \left|h \right|^2 $ is the channel gain of each node without loss of generality.
	\end{itemize}
	
	\subsection{Open Problems}
	
	Neither orthogonal communication nor CoMAC can be implemented directly in the disorganized network. For orthogonal communication, the collection of individual data from nodes results in excessive latency as the number of nodes increases. As for CoMAC, different node has a different transmission destination  in the disorganized network  instead of the same transmission destination in the relay-free network. This implies that it is impossible to use concurrent node transmissions to attain CoMAC. Thus, we expect to expand the analysis of relay-free networks to the analysis of disorganized networks considering the practical case that includes channel fading, noise, and resource allocation. Since the analysis of disorganized networks is more general but also challenging, several issues need to be solved.
	
	\begin{enumerate}[1.]
		\item The disorganized network needs to be reorganized to make further analysis possible. It needs to be considered that how to recast the disorganized network into a hierarchical network in a general way and how to design a scheme that ensures the reliable computation of the desired function at the fusion center.
		\item With the proposed scheme, it becomes important to evaluate the performance through computation rate and further optimize it against channel fading. Thus, the corresponding computation rate should be derived and the resource allocation should be discussed.
	\end{enumerate}

	\section{Reorganization of Disorganized Networks}\label{Reorganization of Disorganized Networks}
	Although the disorganized network in Fig.~\ref{fig:randomnetwork} is general and practical, its structure makes analysis difficult. Before proposing the scheme to efficiently compute the desired function at the fusion center, in Section \ref{Hybrid Aggregation for Single Layer}, we first reorganize the disorganized network into a hierarchical network with multiple layers, which consists of groups and subgroups. In Section \ref{Computing and Communicating Functions}, we present a detailed description of the division and reconstruction of the desired function and the group function.
	
	\begin{figure*}
		\centering
		\subfloat[The structure of one subgroup]{
			\includegraphics[width=0.49\linewidth]{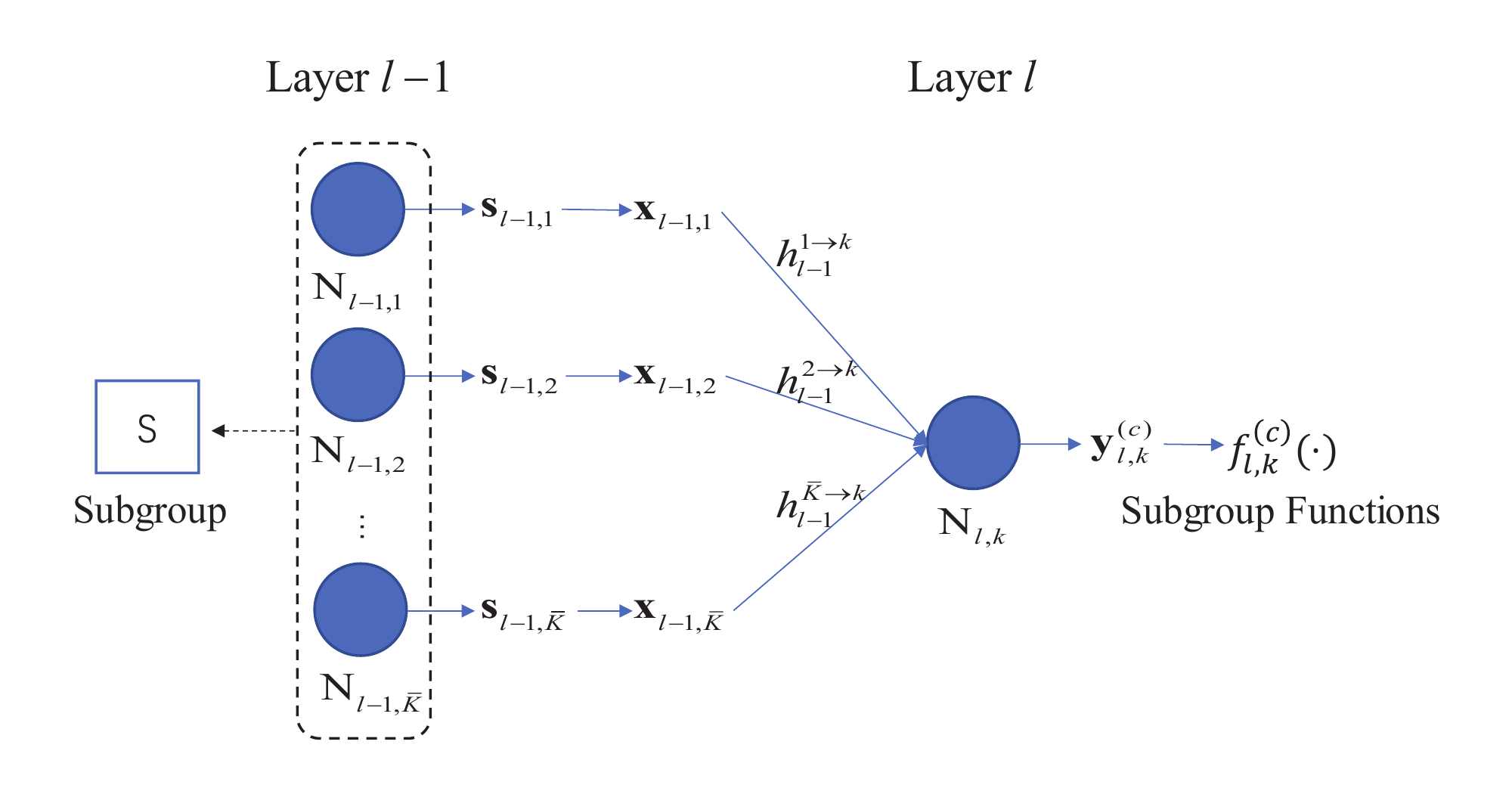}
			\label{fig:subgroup}
		}
		\subfloat[The structure of one group]{
			\includegraphics[width=0.49\linewidth]{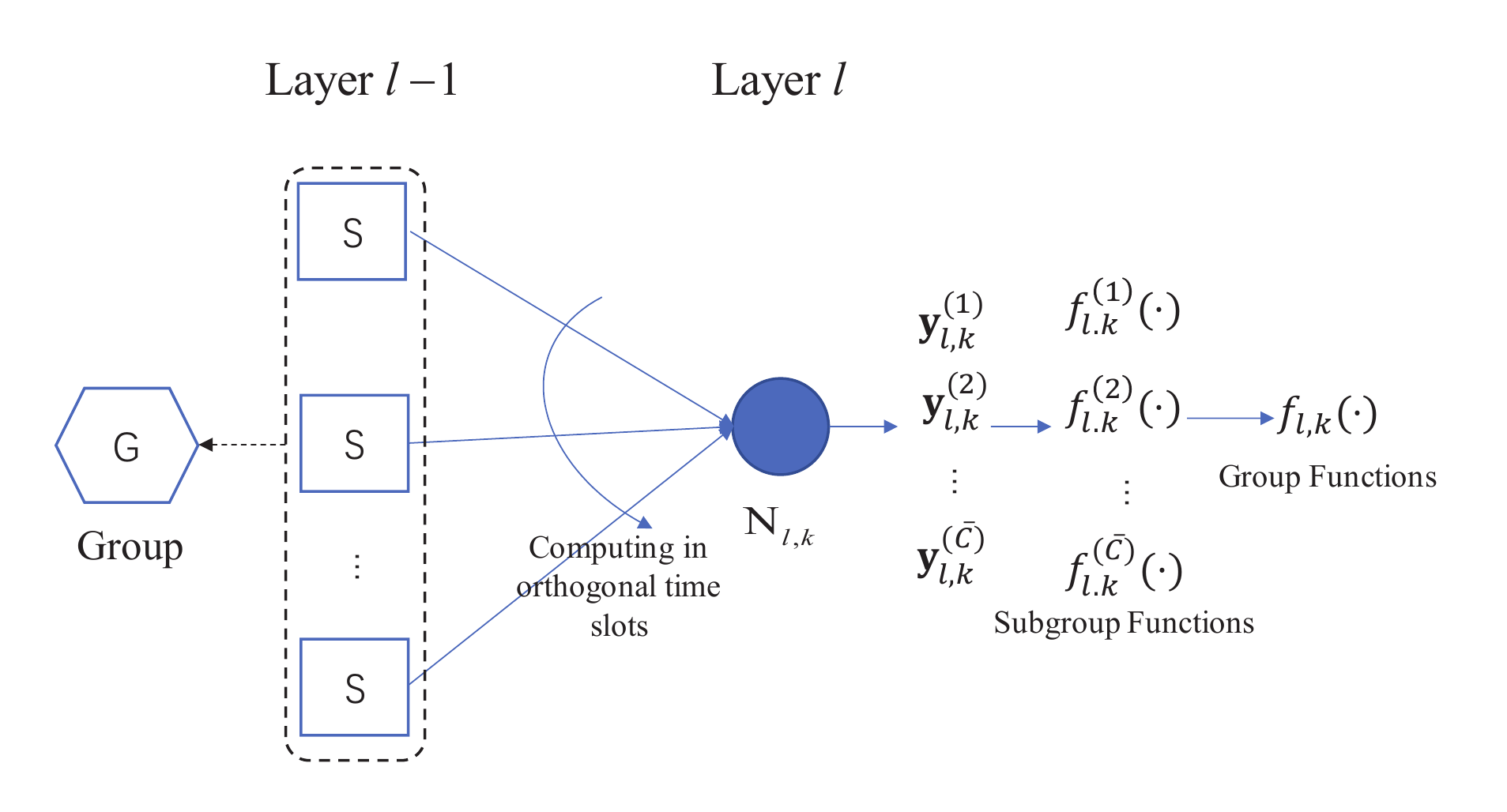}
			\label{fig:group}
		}
		\hfill
		\captionsetup[subfigure]{oneside,margin={0.5cm,0cm}}
		\begin{minipage}{0.48\linewidth}
			\subfloat[From one layer to another layer]{
				\hspace*{0.8cm}
				
				\includegraphics[scale=0.6,valign=c]{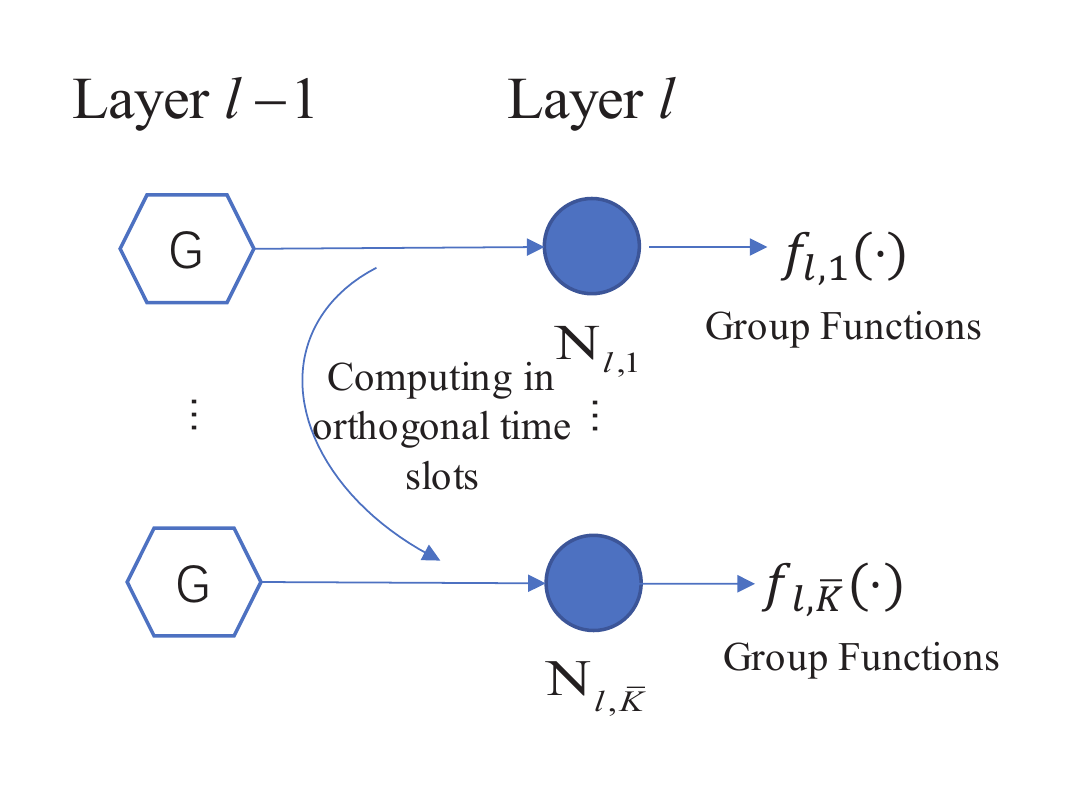}
				\label{fig:LayertoLayer}
			}
		\end{minipage}
		\subfloat[The topology of the hierarchical network]{
			\includegraphics[width=0.48\linewidth,valign=c]{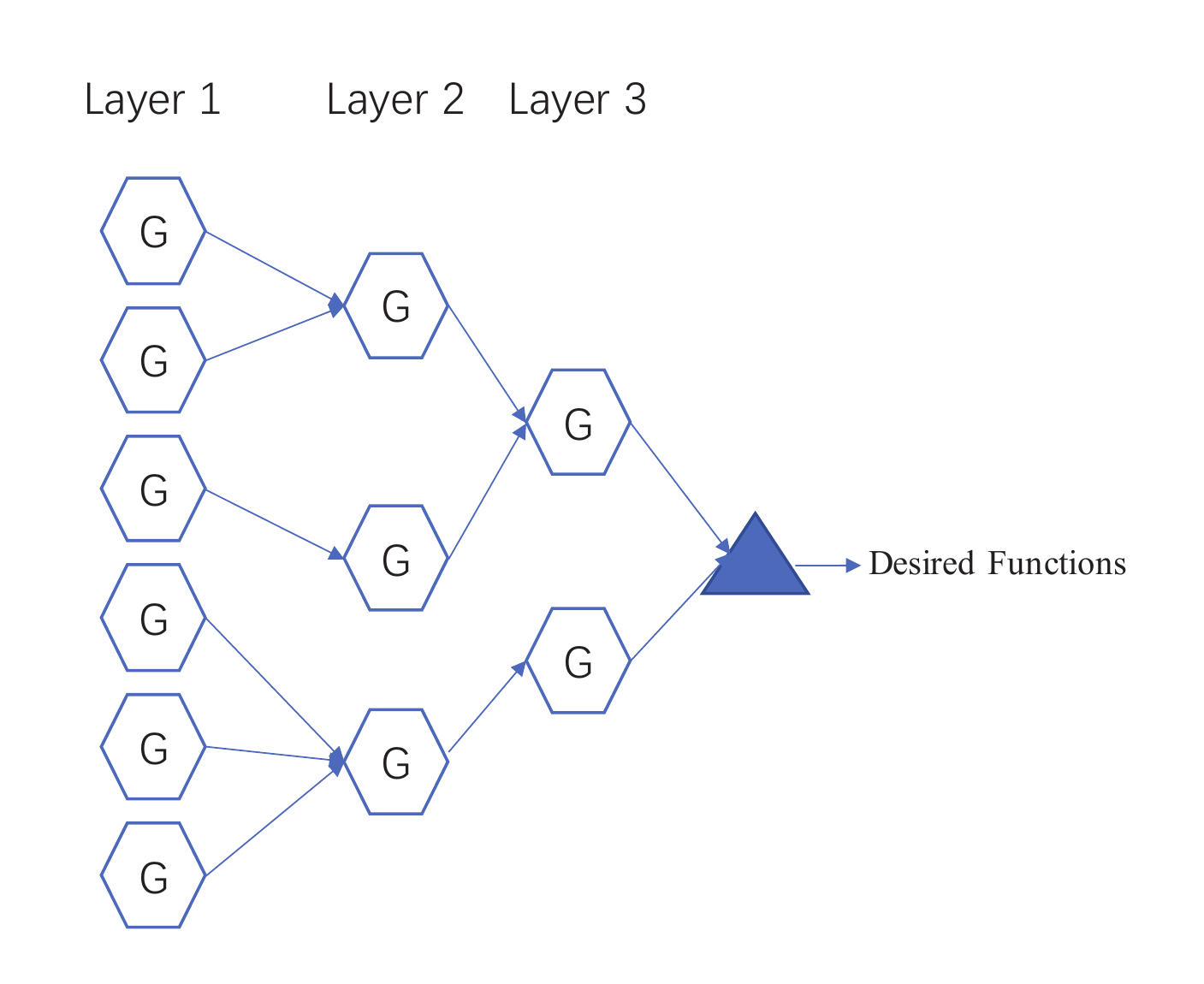}
			\label{fig:multilayer}
		}
		\caption{Reorganization of the disorganized network}
		\label{fig:networkb}
	\end{figure*}
	
	\subsection{Hierarchical Networks}\label{Hybrid Aggregation for Single Layer}\label{Hierarchical Networks}
	
	We introduce two components to reorganize the disorganized network, which are given as follows.
	
	\begin{itemize}
		\item \textbf{\emph{Subgroups.}} 
		Assume there are two layers, namely the $ (l-1) $-th layer and the $ l $-th layer. We place $ \bar{K} $ nodes\footnote{Since the number of nodes in the $ (l-1) $-th layer can be arbitrary, we assume that it is $ \bar{K} $ without loss of generality.} in the $ (l-1) $-th layer and one node in the $ l $-th layer as $ {\rm N}_{l,k} $, where these $ \bar{K} $ nodes wish to transmit their data to $ {\rm N}_{l,k} $. The subgroup, whose index is $ c $, consists of these $ \bar{K} $ nodes where the $ i $-th node is denoted as $ {\rm N}_{l-1,i} $, and the node $ {\rm N}_{l,k} $ computes the subgroup function $ f^{(c)}_{l,k}(\cdot) $ associated with the subgroup $ c $.
		
		As given in Fig.~\ref{fig:subgroup}, each node $ {\rm N}_{l-1,i} $ in the $ (l-1) $-th layer owns a data vector $ \bm{\mathrm{s}}_{l-1,i} $ of length $ T_d $. In the $ l $-th layer, the node $ {\rm N}_{l,k} $  computes the subgroup function $ f^{(c)}_{l,k}(\cdot) $ via concurrent node transmissions. Based on Section \ref{Aggregation Schemes}, CoMAC is applied to compute the subgroup function.
		\item \textbf{\emph{Groups.}} As shown in Fig.~\ref{fig:group}, one group, which is allocated to the node $ {\rm N}_{l,k} $ in the $ l $-th layer, consists of several subgroups in the $ (l-1) $-th layer. Assume the number of subgroups is $ \bar{C} $\footnote{Since the number of subgroups for a group can be arbitrary, we assume that it is $ \bar{C} $ without loss of generality.}, then the node $ {\rm N}_{l,k} $ needs to reconstruct the group function $ f_{l,k}(\cdot) $ using  $ \bar{C} $  subgroup functions.
		
		To reconstruct the group function $f_{l,k}(\cdot) $ at $ {\rm N}_{l,k} $, all the subgroup functions must be obtained first. Based on Section \ref{Aggregation Schemes}, orthogonal communication is applied by communicating the subgroup functions to $ {\rm N}_{l,k} $ during the given channel uses, where different subgroup is active to compute different subgroup function $ f^{(c)}_{l,k}(\cdot) $ at different channel use. After obtaining all the subgroup functions $ \{f^{(c)}_{l,k}(\cdot)\}_{c=1}^{\bar{C}} $, the node $ {\rm N}_{l,k} $ can reconstruct the group function $ f_{l,k}(\cdot) $.
	\end{itemize}
	
	\begin{remark}[Generalization of Classical Aggregation Schemes]
		With the description of two components, one can observe that the classical aggregation schemes in relay-free networks are combined in our scheme, where the subgroup function is obtained by CoMAC whereas the group function is obtained by orthogonal communication. Also, the structure of the group shown as Fig.~\ref{fig:group} is the relay-free network in a general way. By setting the number of subgroups $ \bar{C} $ to one, CoMAC is generalized, which implies that the only one subgroup is treated as a group. Also, by setting the number of subgroups $ \bar{C} $ to the number of nodes $ \bar{K} $, orthogonal communication is generalized, which implies that each node as a subgroup transmits its data individually.
	\end{remark}
	
	With the help of subgroups and groups, the aggregation from the $ (l-1) $-th layer to the $ l $-th layer is given as Fig.~\ref{fig:LayertoLayer}. Each node $ {\rm N}_{l,k} $ in the $ l $-th layer serves a group to reconstruct the corresponding group function $ f_{l,k}(\cdot) $. Similar to orthogonal communication, different group in the $ (l-1) $-th layer is allocated some orthogonal channel uses to compute the subgroup functions and reconstruct the group function at the corresponding node in the $ l $-th layer. Based on Fig.~\ref{fig:LayertoLayer}, we further expand it to the case with multiple layers. Then, the hierarchical network is obtained as shown in Fig.~\ref{fig:multilayer}. 
	
	\textbf{\emph{Hierarchical Networks.}} The disorganized network in Fig.~\ref{fig:randomnetwork} is reorganized into the hierarchical network, which consists of $ L $ $ (L\geq2) $ layers where the $ l $-th layer includes $ K_l $ nodes and the indexes of them belong to a set $ \mathcal{K}_{l} $. Compared with the disorganized network, in the hierarchical network, $ K_{1}$ nodes in the first layer are the source nodes, the nodes from the second layer to the $ (L-1) $-th layer are relay nodes, and the only one node in the $ L $-th layer is regarded as a fusion center. Finally, the desired function will be computed at the fusion center over layers.
	
	\begin{remark}[Equivalent Relation Between Disorganized Networks and Hierarchical Networks]
		In both disorganized networks and hierarchical networks, the routing path of each node does not be changed, which implies that the transmission destination of one node in the disorganized network is the same as the one in the corresponding hierarchical network. Further, the condition, where each node only owns one transmission destination, is satisfied in both disorganized networks and hierarchical networks. Thus, we can always find an equivalent hierarchical network by changing the parameters of the hierarchical network to replace the disorganized one.
	\end{remark}
	
	\subsection{Division and Reconstruction of Functions}\label{Computing and Communicating Functions}
	
	In the hierarchical network including $ L $ layers, the fusion center computes the desired function $ f\left( \left\lbrace s_{1,k}[j]\right\rbrace _{k\in\mathcal{K}_1}\right)  $, where $ s_{1,k}[j] $ is the data sampled by the node $ {\rm N}_{1,k}, k\in\mathcal{K}_1 $. Depending on a given topology, all these nodes in $ \mathcal{K}_{l-1} $ are divided into $ K_{l} $ groups and allocated to $ K_{l} $ nodes in the $ l $-th layer for $ l\ge2 $. We define the set $ \mathcal{K}_{{\rm N}_{l,k}} $ including the indexes of the nodes in the group allocated to $ {\rm N}_{l,k} $. Thus, the desired function is divided into group functions and each group function, associated with the data $ \left\lbrace s_{l-1,i}[j]\right\rbrace_{i\in\mathcal{K}_{\mathrm{N}_{l,k}}} $, is computed at $ {\rm N}_{l,k} $. The detailed definition of the group function is given as follows.
	
	\begin{definition}[Group Function]\label{Group Function}
		For $ l\ge2 $, let 
		\begin{equation}\label{key}
		\mathcal{K}_{\mathrm{N}_{l,k}}=\left\lbrace x :x\in\mathcal{K}_{l-1} \right\rbrace
		\end{equation} 
		denote a set including these indexes of the nodes as a group allocated to $ \mathrm{N}_{l,k} $. Each element $ x $ in $ \mathcal{K}_{\mathrm{N}_{l,k}} $ is the index of a node from the set $\mathcal{K}_{l-1}$. Suppose that $ \bigcup_{k\in\mathcal{K}_{l}}\mathcal{K}_{\mathrm{N}_{l,k}}=\mathcal{K}_{l-1} $ and $ \mathcal{K}_{\mathrm{N}_{l,u}}\bigcap\mathcal{K}_{\mathrm{N}_{l,v}} =\emptyset$ for all $ u,v\in\mathcal{K}_{l} $. A function $ f_{l,k}(\left\lbrace s_{l-1,i}[j]\right\rbrace_{i\in\mathcal{K}_{\mathrm{N}_{l,k}}} ) $ is said to be a group function if and only if there exists a function $ g_{l}(\cdot) $ satisfying
		\begin{equation}\label{key}
		\begin{split}
		f(\bm{\mathrm{s}}_1[j])=g_{l}(&f_{l,1}(\left\lbrace s_{l-1,i}[j]\right\rbrace_{i\in\mathcal{K}_{\mathrm{N}_{l,1}}} ), f_{l,2}(\left\lbrace s_{l-1,i}[j]\right\rbrace_{i\in\mathcal{K}_{\mathrm{N}_{l,2}}} ),\\
		& \cdots, f_{l,K_l }(\left\lbrace s_{l-1,i}[j]\right\rbrace_{i\in\mathcal{K}_{\mathrm{N}_{l,K_l}}} ))
		\end{split}
		\end{equation}
		for $ l\ge2 $.
	\end{definition}
	
	Definition \ref{Group Function} suggests that the group functions in each layer can reconstruct the desired function, even though the desired function only needs to be reconstructed at the fusion center in the last layer by these group functions in the $ (L-1) $-th layer.
	
	To attain the computation of the group function at  $ \mathrm{N}_{l,k} $, all these subgroup functions should be obtained at $ \mathrm{N}_{l,k} $ first since a group function is further divided into several subgroup functions\footnote{Shown in Fig.~\ref{fig:group}, a group is divided into several subgroups.}. The function computed by a subgroup is called a subgroup function, and its definition is given as follows.
	
	\begin{definition}[Subgroup Function]\label{subgroup Function}
		Assume the nodes in $ \mathcal{K}_{{\rm N}_{l,k}} $ as a group is divided into $ C_{\mathrm{N}_{l,k}}  $ subgroups. The set $ \mathcal{C}_{\mathrm{N}_{l,k}} $ includes indexes of these $ C_{\mathrm{N}_{l,k}}  $ subgroups satisfying that $ \bigcup_{c\in\mathcal{C}_{\mathrm{N}_{l,k}}}\mathcal{K}^{(c)}_{\mathrm{N}_{l,k}}=\mathcal{K}_{\mathrm{N}_{l,k}} $, where $ \mathcal{K}^{(c)}_{\mathrm{N}_{l,k}} \subseteq\mathcal{K}_{\mathrm{N}_{l,k}} $ and $ \mathcal{K}^{(u)}_{\mathrm{N}_{l,k}}\bigcap\mathcal{K}^{(v)}_{\mathrm{N}_{l,k}} =\emptyset$ for all $ u,v\in\mathcal{C}_{\mathrm{N}_{l,k}} $. A function $ f^{(c)}_{l,k}(\left\lbrace s_{l-1,i}[j]\right\rbrace_{i\in\mathcal{K}^{(c)}_{\mathrm{N}_{l,k}}} ) $ is said to be a subgroup function if and only if there exists a function $ g_{l,k}(\cdot) $ satisfying 
		\begin{equation}\label{key}
		\begin{split}
		&f_{l,k}(\left\lbrace s_{l-1,i}[j]\right\rbrace_{i\in\mathcal{K}_{\mathrm{N}_{l,k}}})\\
		&=g_{l,k}(f^{(1)}_{l,k}(\left\lbrace s_{l-1,i}[j]\right\rbrace_{i\in\mathcal{K}^{(1)}_{\mathrm{N}_{l,k}}},\cdots,\quad f^{(C_{\mathrm{N}_{l,k}})}_{l,k}(\left\lbrace s_{l-1,i}[j]\right\rbrace_{i\in\mathcal{K}^{(C_{\mathrm{N}_{l,k}})}_{\mathrm{N}_{l,k}}} ) ).		\end{split}
		\end{equation}
	\end{definition}
	
	The property of subgroup functions is similar to the one of group functions, which shows that a group function $ f_{l,k}(\left\lbrace s_{l-1,i}[j]\right\rbrace_{i\in\mathcal{K}_{\mathrm{N}_{l,k}}} ) $ can be reconstructed at  $ \mathrm{N}_{l,k} $ after $ \mathrm{N}_{l,k} $ obtains  $ C_{\mathrm{N}_{l,k}}  $ subgroup functions.
	
	To compute the subgroup functions reliably against noise, we apply sequences of nested lattice codes. For the node $ {\rm N}_{l-1,i} $ in the $ (l-1) $-th layer, the length-$ T_d $ data vector $ \bm{\mathrm{s}}_{l-1,i} $ is mapped to the length-$ \bar{n} $ transmitted vector  $ \bm{\mathrm{x}}_{l-1,i}=[x_{l-1,i}[1],x_{l-1,i}[2],\cdots,x_{l-1,i}[\bar{n}]] $. Then, similar to Eq.~\eqref{model of snetwork}, the length-$ \bar{n} $ received vector $ \bm{\mathrm{y}}^{(c)}_{l,k} $ of the $ c $-th subgroup at $ {\rm N}_{l,k} $ is given as
	\begin{equation}\label{System Model}
	y^{(c)}_{l,k}[m]=\sum_{i\in\mathcal{K}^{(c)}_{\mathrm{N}_{l,k}}}v^{i \to k}_{l-1}[m]h^{i\to k}_{l-1}[m]x^{i \to k}_{l-1}[m]+w[m],
	\end{equation}
	where $ h^{i \to k}_{l-1}[m] $ is the channel from $ \mathrm{N}_{l-1,i} $ to $ \mathrm{N}_{l,k} $ at the $ m $-th channel use, $ x^{i \to k}_{l-1}[m] $ is the $ m $-th element of the transmitted vector $ \bm{\mathrm{x}}_{l-1,i} $, $ v_{l-1}^{i\to k}[m]=\frac{|h_{l-1}^{i\to k}[m]|}{|h_{l-1}^{i\to k}[m]}\sqrt{P_{l-1}^{i\to k}[m]} $ is the power factor of $ \mathrm{N}_{l-1,i} $, $ P_{l-1}^{i\to k}[m] $ is the transmitted power of $ \mathrm{N}_{l-1,i} $ and $ w[m] $ is i.i.d. complex Gaussian random noise following $ \mathcal{CN}(0,1) $.
	
	After receiving $ \bm{\mathrm{y}}^{(c)}_{l,k} $, the decoding function is used to unmap the length-$ \bar{n} $ received vector to $ T_d $ subgroup functions $ \{f^{(c)}_{l,k}(\lbrace s_{l-1,i}[j]\rbrace_{i\in\mathcal{K}^{(c)}_{\mathrm{N}_{l,k}}})\}_{j=1}^{T_d} $.
	
	\begin{figure*}
		\centering
		\includegraphics[width=\linewidth]{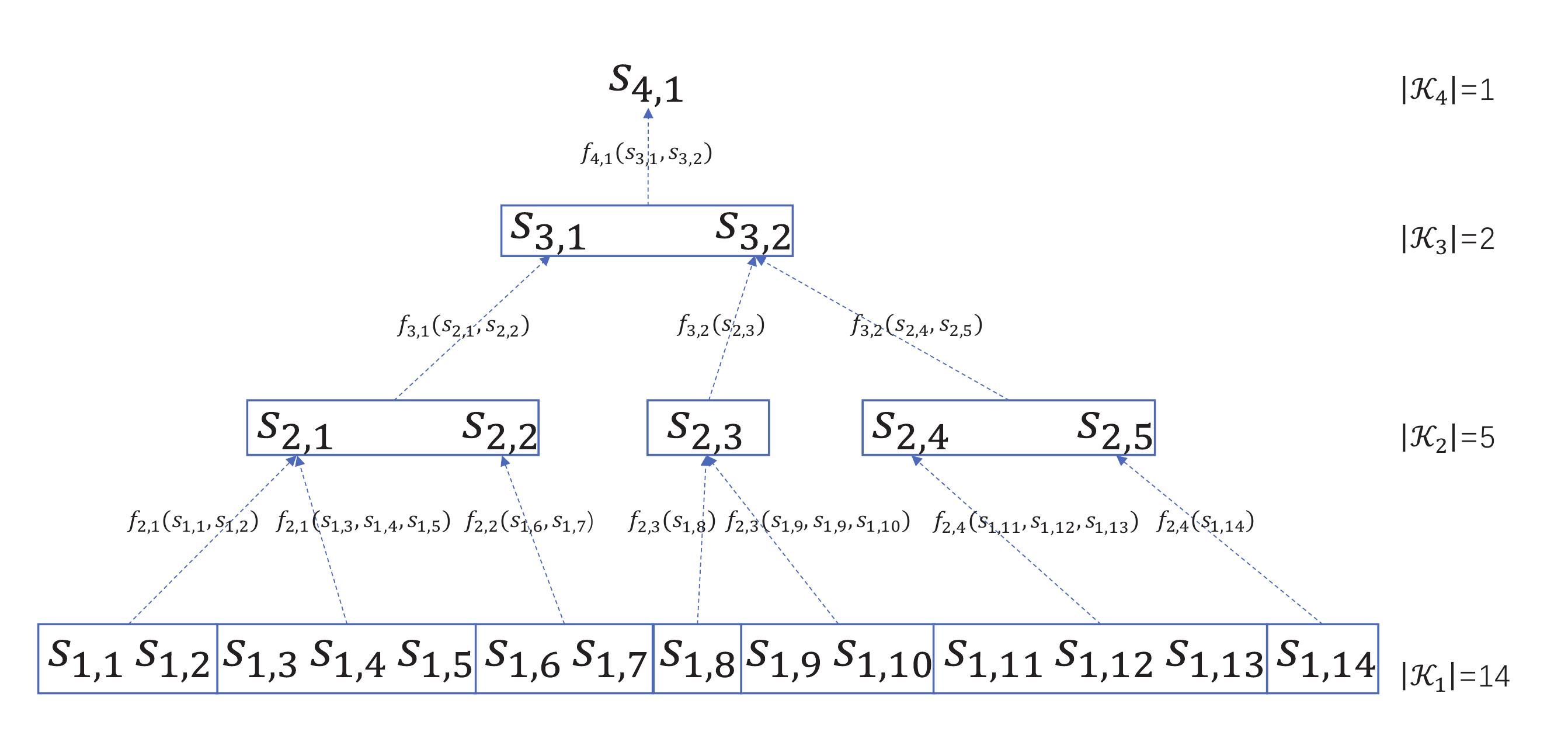}
		\caption{Procedure of ML-FC}
		\label{fig:functionflow}
	\end{figure*}
	
	\begin{algorithm}\label{XXX}
		\caption{Function Computation and Communication Procedure}
		
		\begin{minipage}{0.95\linewidth}
			\begin{enumerate}[1.]
				\item \textbf{Initialization for Source Nodes}:\\
				The sources nodes in the first layer are divided into five groups, each group is allocated to a node in the second layer. The data of $ {\rm N}_{1,k} $ is $ s_{1,k} $, which is drawn from the corresponding random source.
				\item\label{begin1}\label{end1} \textbf{Procedure in One Group}:
				\begin{enumerate}[Step 1.]
					\item The given channel uses for the group $ \left\lbrace {\rm N}_{1,1},{\rm N}_{1,2},{\rm N}_{1,3},{\rm N}_{1,4},{\rm N}_{1,5} \right\rbrace  $ belongs a set $ \mathcal{T}_{2,1} $, and the group function needs to be computed at $ {\rm N}_{2,1}  $ during these channel uses.
					\item To obtain the group function, two subgroup functions should first be computed at $ {\rm N}_{2,1}  $ using CoMAC in different channel uses from $ \mathcal{T}_{2,1} $ since the group consists of two subgroups.
					\item After $ | \mathcal{T}_{2,1} | $ channel uses, at $ {\rm N}_{2,1}  $, all subgroup functions are computed. $ f^{(1)}_{2,1}(s_{1,1},s_{1,2}) $ is the subgroup function associated with the first subgroup and $ f^{(2)}_{2,1}(s_{1,3},s_{1,4},s_{1,5}) $ is the subgroup function associated with the second subgroup.
					\item Using these subgroup functions, the corresponding group function is reconstructed at $ {\rm N}_{2,1}  $ as  $ f_{2,1}(s_{1,1},s_{1,2},s_{1,3},s_{1,4},s_{1,5})=g_{2,1}(f^{(1)}_{2,1}(s_{1,1},s_{1,2}),f^{(2)}_{2,1}(s_{1,3},s_{1,4},s_{1,5})) $.
				\end{enumerate}
				\item \textbf{From One Layer to Another Layer}:\\
				Using the same steps in the procedure in one group, each group finishes the computation of the group function at the corresponding node in the second layer.
				\item \textbf{Initialization for Relay Nodes}:\\
				In the second layer, the data of $ {\rm N}_{2,k} $ is $ s_{2,k} $ satisfying $ s_{2,k}=f_{2,k}(\cdot) $. All the nodes in $ \mathcal{K}_2 $ is divided into two groups that are allocated to the nodes in the third layer.
				
				\item \textbf{Reconstruction of Desired Function}:\\
				Using similar steps in the procedure in one group from one layer to another layer, finally, the desired function is obtained as $ f(\{s_{1,i}\}_{i\in\mathcal{K}_1})=g_{4,1}(f_{4,1}(s_{3,1},s_{3,2})) $ at the fusion center.
			\end{enumerate}
		\end{minipage}
		
	\end{algorithm}

	\section{Computation Rates in Hierarchical Networks}\label{Computation Rate of Multi-layer Networks}
	
	With the help of the hierarchical network, the analysis of the disorganized network becomes tractable. First of all, we provide a detailed procedure of ML-FC in Section \ref{Function Computation Procedure}. The procedure helps us to rule the relation of the functions between nodes. Then, in Section \ref{Achievable Computation Rate}, we derive the computation rate of ML-FC based on the relation of the functions.
	
	\subsection{Procedure of ML-FC}\label{Function Computation Procedure}
	Fig.~\ref{fig:functionflow} is an example of the hierarchical network with the given topology, which aims at computing the desired function $ f\left( \left\lbrace s_{1,i}\right\rbrace _{i\in\mathcal{K}_1}\right)  $ over $ 4 $ layers. For $ l\ge2 $, each node $ {\rm N}_{l,k} $ is assigned a group consisting of several subgroups. The data of $ {\rm N}_{l,k} $ is denoted as $ s_{l,k} $. We describe the procedure of ML-FC as Algorithm \ref{XXX}.
	
	With the help of the example shown in Fig.~\ref{fig:functionflow}, we extend it to a general case mentioned in Section \ref{Computing and Communicating Functions} and show the recurrence relation between these functions over layers.
	
	\begin{remark}[Relation Between Functions in Hierarchical Networks]
		In the second layer, each node $ {\rm N}_{2,k} $ computes $ C_{{\rm N}_{2,k}}  $ subgroup functions. Then, it reconstructs the group function as
		\begin{equation}\label{key}
		\begin{split}
		&f_{2,k}\left(\left\lbrace\left\lbrace s_{1,i}\right\rbrace_{i\in\mathcal{K}^{(c)}_{{\rm N}_{2,k}}}\right\rbrace_{c\in\mathcal{C}_{{\rm N}_{2,k}}}\right)=g_{2,k}\left(\left\lbrace f^{(c)}_{2,k} \left(\left\lbrace s_{1,i}\right\rbrace_{i\in\mathcal{K}^{(c)}_{{\rm N}_{2,k}}} \right)\right\rbrace_{c\in\mathcal{C}_{{\rm N}_{2,k}}}   \right).
		\end{split}
		\end{equation}
		
		By setting the data $ s_{2,k}=f_{2,k}(\lbrace\lbrace s_{1,i}\rbrace_{i\in\mathcal{K}^{(c)}_{{\rm N}_{2,k}}}\rbrace_{c\in\mathcal{C}_{{\rm N}_{2,k}}}) $ for each node in $ \mathcal{K}_2 $, the node $ {\rm N}_{3,k} $ in the third layer also reconstructs the corresponding group function as
		\begin{equation}\label{key}
		\begin{split}
		&f_{3,k}\left(\left\lbrace\left\lbrace s_{2,i}\right\rbrace_{i\in\mathcal{K}^{(c)}_{{\rm N}_{3,k}}}\right\rbrace_{c\in\mathcal{C}_{{\rm N}_{3,k}}}\right)=g_{3,k}\left(\left\lbrace f^{(c)}_{3,k} \left(\left\lbrace s_{2,i}\right\rbrace_{i\in\mathcal{K}^{(c)}_{{\rm N}_{3,k}}} \right)\right\rbrace_{c\in\mathcal{C}_{{\rm N}_{3,k}}}   \right).
		\end{split}
		\end{equation}
		
		Thus, we can obtain the recurrence relation between the $ l $-th layer and the $ (l-1) $-th layer as
		\begin{equation}\label{the recurrence relation}
		\begin{split}
		&f_{l,k}\left(\left\lbrace\left\lbrace s_{l-1,i}\right\rbrace_{i\in\mathcal{K}^{(c)}_{{\rm N}_{l,k}}}\right\rbrace_{c\in\mathcal{C}_{{\rm N}_{l,k}}}\right)=g_{l,k}\left(\left\lbrace f^{(c)}_{l,k} \left(\left\lbrace s_{l-1,i}\right\rbrace_{i\in\mathcal{K}^{(c)}_{{\rm N}_{l,k}}} \right)\right\rbrace_{c\in\mathcal{C}_{{\rm N}_{l-1,k}}}   \right),
		\end{split}
		\end{equation}
		where $ s_{l,k}=f_{l,k}(\lbrace\lbrace s_{l-1,i}\rbrace_{i\in\mathcal{K}^{(c)}_{{\rm N}_{l,k}}}\rbrace_{c\in\mathcal{C}_{{\rm N}_{l,k}}}) $.
		
		At the last layer, i.e., $ l=L $, only including the fusion center, the desired function is finally computed because of
		\begin{equation}\label{Reliable Desired Function}
		\begin{split}
		f_{L}\left(\left\lbrace\left\lbrace s_{L-1,i}\right\rbrace_{i\in\mathcal{K}^{(c)}_{{\rm N}_{L}}}\right\rbrace_{c\in\mathcal{C}_{{\rm N}_{L}}}\right)&\stackrel{(a)}{=}f_{L}\left(\left\lbrace s_{L-1,i}\right\rbrace_{i\in\mathcal{K}_{L-1}}\right)\\
		&\stackrel{(b)}{=}f_{L}\left(\left\lbrace s_{L-2,i}\right\rbrace_{i\in\mathcal{K}_{L-2}}\right)\\
		&=\cdots\\
		&\stackrel{(c)}{=}f_{L}\left(\left\lbrace s_{1,i}\right\rbrace_{i\in\mathcal{K}_{1}}\right),
		\end{split}
		\end{equation}
		where the condition $ (a) $ follows since $ K_L=1 $, the condition $ (b) $ follows because the values of $ \left\lbrace s_{L-1,i}\right\rbrace_{i\in\mathcal{K}_{L-1}} $ are associated with $ \left\lbrace s_{L-2,i}\right\rbrace_{i\in\mathcal{K}_{L-2}} $ and the condition $ (c) $ follows due to the recurrence relation (Eq.~\eqref{the recurrence relation}). 
	\end{remark}
	
	\subsection{Achievable Computation Rates}\label{Achievable Computation Rate}
	
	Eq.~\eqref{the recurrence relation} shows that the computation rate of the desired function is determined by all the group functions over $ L $ layers, and each group function is reconstructed by the corresponding subgroup functions. Thus, we present the computation rates of the subgroup function, the group function and the desired function step by step.
	
	\begin{lemma}[Rate of Subgroup Function]\label{Rate of Subgroup Function}
		For a subgroup $ \mathcal{K}^{(c)}_{{\rm N}_{l,k}} $ with $ K^{(c)}_{{\rm N}_{l,k}} $ nodes, the computation rate of the subgroup function $ f^{(c)}_{l,k} (\lbrace s_{l-1,i}\rbrace_{i\in\mathcal{K}^{(c)}_{{\rm N}_{l,k}}} ) $ at the $ m $-th channel use is given as
		\begin{equation}\label{key}
		R_{l,k}^{(c)}[m]=\mathsf{C}^{+}\left(\dfrac{1}{ K^{(c)}_{{\rm N}_{l,k}}}+\min_{i\in \mathcal{K}^{(c)}_{{\rm N}_{l,k}}}\left[|h^{i\to k}_{l-1}[m]|^2P^{i\to k}_{l-1}[m]\right]\right),
		\end{equation}
		where  $ |h^{i \to k}_{l-1}[m]|^2 $ is the channel gain and $ P_{l-1}^{i\to k}[m] $ is the transmitted power (see Eq.~\eqref{System Model}).
		\begin{IEEEproof}
			Please refer to Eq.~\eqref{System Model} and  \cite[Theorem 3 and Section IV-A]{jeon2014computation}.
		\end{IEEEproof}
	\end{lemma}
	
	To reconstruct the group function $ f_{l,k}(\lbrace\lbrace s_{l-1,i}\rbrace_{i\in\mathcal{K}^{(c)}_{{\rm N}_{l,k}}}\rbrace_{c\in\mathcal{C}_{{\rm N}_{l,k}}}) $ during the given $ | \mathcal{T}_{l,k} | $ channel uses where the set $ \mathcal{T}_{l,k} $ includes the channel uses for $ {\rm N}_{l,k} $, the subgroup functions should be computed first at different channel use from $ \mathcal{T}_{l,k} $. Assume that the channel uses allocated to the corresponding subgroup $ \mathcal{K}^{(c)}_{{\rm N}_{l,k}} $ are in a set $ \mathcal{T}^{(c)}_{l,k} \subseteq \mathcal{T}_{l,k} $ satisfying $ | \mathcal{T}^{(c)}_{l,k} |=\beta_{l,k}^{(c)}| \mathcal{T}_{l,k} | $ and $ \sum_{c\in\mathcal{C}_{{\rm N}_{l,k}}}\beta_{l,k}^{(c)}=1 $. After obtaining all the subgroup functions, the group function is reconstructed by Eq.~\eqref{the recurrence relation}. Thus, the computation rate of the group function is given as follows.
	
	\begin{theorem}[Rate of Group Function]\label{General Rate of Minimal Substructure}
		For any group $ \mathcal{K}_{{\rm N}_{l,k}} $ with $ C_{{\rm N}_{l,k}} $ subgroups, the computation rate of the group function reconstructed at $ {\rm N}_{l,k} $ is
		\begin{equation}
			\begin{split}
			R_{l,k}=&\min_{c\in\mathcal{C}_{{\rm N}_{l,k}}}\frac{\beta^{(c)}_{l,k}}{| \mathcal{T}^{(c)}_{l,k} |}\sum_{m\in\mathcal{T}^{(c)}_{l,k}}\left[\mathsf{C}^{+}\left(\dfrac{1}{K^{(c)}_{{\rm N}_{l,k}}}+\min_{i\in \mathcal{K}^{(c)}_{{\rm N}_{l,k}}}\left[|h^{i\to k}_{l-1}[m]|^2P^{i\to k}_{l-1}[m]\right]\right)\right]\\
			=&\min_{c\in\mathcal{C}_{{\rm N}_{l,k}}} \beta^{(c)}_{l,k} \mathsf{E}\left[\mathsf{C}^{+}\left(\dfrac{1}{K^{(c)}_{{\rm N}_{l,k}}}+\min_{i\in \mathcal{K}^{(c)}_{{\rm N}_{l,k}}}\left[|h^{i\to k}_{l-1}|^2P^{i\to k}_{l-1}\right]\right) \right].
			\end{split}
		\end{equation}
	\end{theorem}
	\begin{IEEEproof}
		Based on Lemma \ref{Rate of Subgroup Function}, the average computation rate
		\begin{equation}\label{key}
		\begin{split}
		R_{l,k}^{(c)}=\frac{1}{| \mathcal{T}_{l,k}^{(c)}|}\sum_{m\in\mathcal{T}^{(c)}_{l,k}}R_{l,k}^{(c)}[m]
		\end{split}
		\end{equation}
		is achievable for computing the subgroup function $ f^{(c)}_{l,k}(\lbrace s_{l-1,i}[j]\rbrace_{i\in\mathcal{K}^{(c)}_{\mathrm{N}_{l,k}}}) $ during $ | \mathcal{T}_{l,k}^{(c)}| $ channel uses when $ | \mathcal{T}_{l,k} | $ increases. Depending on Definition \ref{Computation rate}, the number of the values of the subgroup function computed during $ | \mathcal{T}_{l,k}^{(c)}| $ channel uses is $ U^{(c)}_{l,k}=\frac{R_{l,k}^{(c)}\left| \mathcal{T}_{l,k}^{(c)}\right|}{H(f(\bm{\mathrm{b_v}}))} $. From Eq.~\eqref{the recurrence relation}, we can observe that the group function is reconstructed by $ C_{{\rm N}_{l,k}} $ subgroup functions, which implies that the computation rate of the group function is determined by the rates of these subgroup functions. Since the number of the values of each subgroup function $  U^{(c)}_{l,k} $ is different, only $U_{l,k}= \min_{c\in\mathcal{C}_{{\rm N}_{l,k}}}U^{(c)}_{l,k} $ group functions can be reconstructed. Hence, the computation rate based on Definition \ref{Computation rate} to compute the group function $ f_{l,k}(\lbrace\lbrace s_{l-1,i}\rbrace_{i\in\mathcal{K}^{(c)}_{{\rm N}_{l,k}}}\rbrace_{c\in\mathcal{C}_{{\rm N}_{l,k}}}) $ is
		\begin{equation}\label{key}
		\begin{split}
		R_{l,k}=&\lim\limits_{n\to\infty}\dfrac{U_{l,k}}{| \mathcal{T}_{l,k}|}{H(f(\bm{\mathrm{b_v}}))}\\
		\stackrel{(a)}{=}&\lim\limits_{n\to\infty}\dfrac{\min_{c\in\mathcal{C}_{{\rm N}_{l,k}}}U^{(c)}_{l,k}}{| \mathcal{T}_{l,k}|}{H(f(\bm{\mathrm{b_v}}))}\\
		\stackrel{(b)}{=}&\lim\limits_{n\to\infty}\min_{c\in\mathcal{C}_{{\rm N}_{l,k}}}\frac{R_{l,k}^{(c)}\left| \mathcal{T}_{l,k}^{(c)}\right|}{| \mathcal{T}_{l,k}|}\\
		\stackrel{(c)}{=}&\min_{c\in\mathcal{C}_{{\rm N}_{l,k}}}\beta^{(c)}_{l,k}\mathsf{E}\left[R_{l,k}^{(c)} \right],
		\end{split}
		\end{equation}
		where the condition $ (a) $ follows because of  $U_{l,k}= \min_{c\in\mathcal{C}_{{\rm N}_{l,k}}}U^{(c)}_{l,k} $, the condition $ (b) $ follows because the expression of $ U^{(c)}_{l,k} $ and the condition $ (c) $ follows due to $ \frac{\left| \mathcal{T}_{l,k}^{(c)}\right|}{\left| \mathcal{T}_{l,k}\right|}=\beta_{l,k}^{(c)} $.
	\end{IEEEproof}
	
	To reconstruct the desired function $ f\left( \left\lbrace s_{1,k}\right\rbrace _{k\in\mathcal{K}_1}\right)  $ computed at the fusion center during $ n $ channel uses over $ L $ layers, the group allocated to $ {\rm N}_{l,k} $ is active to compute the group function in the given channel uses in a set $ \mathcal{T}_{l,k} $. Assume the number of the given channel uses is given as $ | \mathcal{T}_{l,k} |=\alpha_{l,k}n $ satisfying $ \sum_{l=2}^{L}\sum_{k\in\mathcal{K}_l}\alpha_{l,k}=1 $. With the help of Theorem \ref{General Rate of Minimal Substructure}, the computation rate of the desired function in the hierarchical network with $ L $ layers is given as follows.
	
	\begin{theorem}[General Rate of Desired Function]\label{General Rate of Multi-Layer Network}
		For any $ L\in\mathbb{N} $ satisfying $ L\ge2 $, the computation rate of the desired function in the hierarchical network over fading MAC is given as
		\begin{equation}
		\begin{split}
		R=&\min_{l\in[2:L]}\min_{k\in\mathcal{K}_{l}}\alpha_{l,k}\min_{c\in\mathcal{C}_{{\rm N}_{l,k}}} \beta^{(c)}_{l,k}\frac{1}{|\mathcal{T}^{(c)}_{l,k}| }\sum_{m\in\mathcal{T}^{(c)}_{l,k}}\left[\mathsf{C}^{+}\left(\dfrac{1}{K^{(c)}_{{\rm N}_{l,k}}}+\min_{i\in \mathcal{K}^{(c)}_{{\rm N}_{l,k}}}\left[|h^{i\to k}_{l-1}[m]|^2P^{i\to k}_{l-1}[m]\right]\right)\right]\\
		=&\min_{l\in[2:L]}\min_{k\in\mathcal{K}_{l}}\alpha_{l,k}\min_{c\in\mathcal{C}_{{\rm N}_{l,k}}} \beta^{(c)}_{l,k}\mathsf{E}\left[\mathsf{C}^{+}\left(\dfrac{1}{K^{(c)}_{{\rm N}_{l,k}}}+\min_{i\in \mathcal{K}^{(c)}_{{\rm N}_{l,k}}}\left[|h^{i\to k}_{l-1}|^2P^{i\to k}_{l-1}\right]\right)\right].
		\end{split}
		\end{equation}
	\end{theorem}
	\begin{IEEEproof}
		Theorem \ref{General Rate of Minimal Substructure} suggests that the computation rate of the group function computed at $ {\rm N}_{l,k} $ is $ R_{l,k} $. However, to reconstruct the group function at $ {\rm N}_{l,k} $, all the nodes in $ \mathcal{K}_{{\rm N}_{l,k}} $ need to obtain the data vector first. In the hierarchical network with $ L $ layers, the data vector of $ {\rm N}_{l-1,i}, i \in \mathcal{K}_{{\rm N}_{l,k}} $ is obtained by the values of the group function computed by the group $ \mathcal{K}_{{\rm N}_{l-1,i}} $ (see Eq.~\eqref{the recurrence relation}). Thus, when considering the relation between layers, the number of the values of the group function computed at $ {\rm N}_{l,k} $ is determined by not only $ \mathcal{K}_{{\rm N}_{l,k}} $ but also $ \{\mathcal{K}_{{\rm N}_{l-1,i}}\}_{i\in\mathcal{K}_{{\rm N}_{l,k}}} $, which is expressed as
		\begin{equation}\label{the recurrence relation again}
		\bar{U}_{l,k}=\min\left\lbrace \dfrac{R_{l,k}| \mathcal{T}_{l,k}|}{H(f(\bm{\mathrm{b_v}}))}, \min_{i\in \mathcal{K}_{{\rm N}_{l,k}}}  \bar{U}_{l-1,i} \right\rbrace.
		\end{equation}
		For the sake of simplicity, we denote $ \frac{R_{l,k}| \mathcal{T}_{l,k}|}{H(f(\bm{\mathrm{b_v}}))} $ as $ \rho_{l,k} $. Based on the recurrence relation (Eq.~\eqref{the recurrence relation again}), at the fusion center ($ l=L $), the number of the values of the desired function is
		\begin{equation}\label{key}
			\begin{split}
			\bar{U}_{L,1}=&\min\left\lbrace \rho_{L,1}, \min_{i\in \mathcal{K}_{{\rm N}_{L,1}}} \bar{U}_{l-1,i} \right\rbrace\\
			\stackrel{(a)}{=}&\min\left\lbrace \rho_{L,1}, \min_{i_1\in \mathcal{K}_{L-1}} \min\left\lbrace \rho_{L-1,i_1}, \min_{i_2\in \mathcal{K}_{{\rm N}_{L-1,i_1}}}  \bar{U}_{L-2,i_2} \right\rbrace \right\rbrace\\
			\stackrel{(b)}{=}&\min\left\lbrace \rho_{L,1},\min_{i_1\in \mathcal{K}_{L-1}} \rho_{L-1,i_1}, \min_{i_1\in \mathcal{K}_{L-1}} \min_{i_2\in \mathcal{K}_{{\rm N}_{L-1,i_1}}}  \bar{U}_{L-2,i_2}\right\rbrace\\
			\stackrel{(c)}{=}&\min\left\lbrace \rho_{L,1},\min_{i_1\in \mathcal{K}_{L-1}} \rho_{L-1,i_1}, \min_{i_1\in \mathcal{K}_{L-2}}  \bar{U}_{L-2,i_2}\right\rbrace\\
			=&\min_{l\in[2:L]}\min_{k\in \mathcal{K}_{l}}\rho_{l,k},
			\end{split}
		\end{equation}
		where the condition $ (a) $ follows because of $ K_L=1 $ and Eq.~\eqref{the recurrence relation again}, the condition $ (b) $ follows since $ \min $ operation is associative and the condition $ (c) $ follows due to $ \cup_{i_1\in\mathcal{K}_{L-1}} \mathcal{K}_{{\rm N}_{L-1,i_1}}=\mathcal{K}_{L-2} $ (see Definition \ref{Group Function}).
		
		At last, the fusion center computes $ \bar{U}_{L,1} $ desired functions  over $ L $ layers. And, the computation rate of the desired function in the hierarchical network is given as
		\begin{equation}\label{key}
		\begin{split}
		R=&\lim\limits_{n\to\infty}\dfrac{\bar{U}_{L,1}}{n}{H(f(\bm{\mathrm{b_v}}))}\\
		\stackrel{(a)}{=}&\lim\limits_{n\to\infty}\dfrac{\min_{l\in[2:L]}\min_{k\in \mathcal{K}_{l}}\rho_{l,k}}{n}{H(f(\bm{\mathrm{b_v}}))}\\
		\stackrel{(b)}{=}&\lim\limits_{n\to\infty}\min_{l\in[2:L]}\min_{k\in \mathcal{K}_{l}}\dfrac{R_{l,k}| \mathcal{T}_{l,k}|}{n}\\
		\stackrel{(c)}{=}&\min_{l\in[2:L]}\min_{k\in\mathcal{K}_{l}}\alpha_{l,k}R_{l,k},
		\end{split}
		\end{equation}
		where the condition $ (a) $ follows because of Eq.~\eqref{the recurrence relation again}, the condition $ (b) $ follows due to $\rho_{l,k}= \frac{R_{l,k}| \mathcal{T}_{l,k}|}{H(f(\bm{\mathrm{b_v}}))} $ and the condition $ (c) $ follows as $ | \mathcal{T}_{l,k} |=\alpha_{l,k}n $.
	\end{IEEEproof}
	
	The rate of Theorem \ref{General Rate of Multi-Layer Network} considers the general case and can reduce to the rate in the relay-free network by setting $ L=2 $. Based on the general rate, we can apply different resource allocation to analyze the corresponding rate and to improve the performance.

	\section{Optimal Resource Allocation}\label{Power Control and Time Allocatio}
	Theorem \ref{General Rate of Multi-Layer Network} suggests that the subgroup with the worst computation rate plays an important role in the hierarchical network. Thus, we consider time allocation and power control in this section to improve the computation rate.
	
	\subsection{Optimal Time Allocation and Fixed Power Control}\label{Fixed Power}
	
	Considering the fixed power constraint for each user, we obtain the computation rate  from Theorem \ref{General Rate of Multi-Layer Network} easily as
	\begin{equation}\label{key}
	\begin{split}
	R=&\min_{l\in[2:L]}\min_{k\in\mathcal{K}_{l}}\alpha_{l,k}\min_{c\in\mathcal{C}_{{\rm N}_{l,k}}} \beta^{(c)}_{l,k}\frac{1}{|\mathcal{T}^{(c)}_{l,k}| }\sum_{m\in\mathcal{T}^{(c)}_{l,k}}\left[\mathsf{C}^{+}\left(\dfrac{1}{K^{(c)}_{{\rm N}_{l,k}}}+\min_{i\in \mathcal{K}^{(c)}_{{\rm N}_{l,k}}}|h^{i\to k}_{l-1}[m]|^2P\right)\right]\\
	\stackrel{(a)}{\le}&\min_{l\in[2:L]}\min_{k\in\mathcal{K}_{l}}\alpha_{l,k}\min_{c\in\mathcal{C}_{{\rm N}_{l,k}}} \beta^{(c)}_{l,k}\mathsf{C}^{+}\left(\dfrac{1}{K^{(c)}_{{\rm N}_{l,k}}}+\mathsf{E}\left[\min_{i\in \mathcal{K}^{(c)}_{{\rm N}_{l,k}}}|h^{i\to k}_{l-1}|^2\right]P\right)
	\end{split}
	\end{equation}
	by setting $ P^{i\to k}_{l-1}[m]=P $, where the condition $ (a) $ follows because of the increase in $ n $ and Jensen's inequality.
	
	One can observe that each $ \alpha_{l,k} $ and each $ \beta^{(c)}_{l,k} $ should be optimized to approach the optimal computation rate since the computation rate of each subgroup function is different. A subgroup function with higher computation rate should be allocated fewer channel uses as the number of the desired functions computed at the fusion center is determined by the minimum of the number of each subgroup function. Thus, we formulate the following optimization problem.
	\begin{problem}
		\begin{align}
		\mathop{\mathrm{maximize}}\limits_{\alpha_{l,k},\beta^{(c)}_{l,k}} & \quad \min_{l\in[2:L]}\min_{k\in\mathcal{K}_{l}}\alpha_{l,k}\min_{c\in\mathcal{C}_{{\rm N}_{l,k}}} \beta^{(c)}_{l,k}\mathsf{C}^{+}\left(\dfrac{1}{K^{(c)}_{{\rm N}_{l,k}}}+ \quad\mathsf{E}\left[\min_{i\in \mathcal{K}^{(c)}_{{\rm N}_{l,k}}}|h^{i\to k}_{l-1}|^2\right]P\right) \nonumber \\
		{\rm s.t.}&\quad  \sum_{l=2}^{L}\sum_{k\in\mathcal{K}_l}\alpha_{l,k}=1 \label{Fixed Power Pro 1 St1}\\
		&\quad  \sum_{c\in\mathcal{C}_{{\rm N}_{l,k}}}\beta^{(c)}_{l,k}=1,\forall l\in[2:L], \forall k\in\mathcal{K}_l \label{Fixed Power Pro 1 St2}
		\end{align}
	\end{problem}
	
	Although the objective function is non-convex, it can be transformed into a convex function by relaxing the parameters through McCormick relaxation  \cite{mitsos2009mccormick} in terms of the bi-linear function. We introduce $ p_{l,k}^{(c)} = \alpha_{l,k}\beta_{l,k}^{(c)}$. Then the constrains \eqref{Fixed Power Pro 1 St1} and \eqref{Fixed Power Pro 1 St2} can be jointly rewritten as $ \sum_{l=2}^{L}\sum_{k\in\mathcal{K}_l}\sum_{c\in\mathcal{C}_{{\rm N}_{l,k}}}p_{l,k}^{(c)}=1 $. Therefore, the $ \max-\min $ problem can be reformed as
	\begin{problem}\label{rewritten max-min}\label{rewritten max-min part one}
		\begin{align}
		\mathop{\mathrm{maximize}}\limits_{p^{(c)}_{l,k},t} & \quad t  \nonumber \\
		{\rm s.t.}&\ p^{(c)}_{l,k}\mathsf{C}^{+}\left(\dfrac{1}{K^{(c)}_{{\rm N}_{l,k}}}+\mathsf{E}\left[\min_{i\in \mathcal{K}^{(c)}_{{\rm N}_{l,k}}}|h^{i\to k}_{l-1}|^2\right]P\right)\ge t,\nonumber\\
		&\quad\quad\quad\quad\quad\forall l\in[2:L],\forall k \in\mathcal{K}_{l}, \forall c \in \mathcal{C}_{l,k}\label{noconvexieq_sad}\\
		&\quad  \sum_{l=2}^{L}\sum_{k\in\mathcal{K}_l}\sum_{c\in\mathcal{C}_{{\rm N}_{l,k}}}p_{l,k}^{(c)}=1 \nonumber
		\end{align}
	\end{problem}
	
	Since Problem \ref{rewritten max-min part one} is a linear programming problem, the problem can be solved by the interior-point methods or Lagrangian duality approach  \cite{boyd2004convex}. However, such an optimal solution requires iteratively updating Lagrange multipliers using sub-gradient methods. By exploring the special structure of Problem \ref{rewritten max-min part one}, we obtain a simple optimal solution that does not require iterations. The optimal $ \left\lbrace  {{p^{*}}^{(c)}_{l,k}}\right\rbrace_{c\in\mathcal{C}_{{\rm N}_{l,k}}}$ and $t^{*} $ can be obtained as closed-form expressions though the Lagrangian function
	\begin{equation}\label{key}
	\begin{split}
	\mathcal{L}=t&-\sum_{l=2}^{L}\sum_{k\in\mathcal{K}_l}\sum_{c\in\mathcal{C}_{{\rm N}_{l,k}}}\lambda_{l,k}^{(c)}\left[ t-p^{(c)}_{l,k}\mathsf{C}^{+}\left(\dfrac{1}{K^{(c)}_{{\rm N}_{l,k}}}+\mathsf{E}\left[\min_{i\in \mathcal{K}^{(c)}_{{\rm N}_{l,k}}}|h^{i\to k}_{l-1}|^2\right]P\right)\right]\\ &-\mu(\sum_{l=2}^{L}\sum_{k\in\mathcal{K}_l}\sum_{c\in\mathcal{C}_{{\rm N}_{l,k}}}p^{(c)}_{l,k}-1),
	\end{split}
	\end{equation}
	where $ \left\lbrace\lambda_{l,k}^{(c)} \right\rbrace  $ and $ \mu $ are Lagrange multipliers.
	
	By setting the first derivative of $ \mathcal{L} $ with respect to $ t $, we have $ \sum_{l=2}^{L}\sum_{k\in\mathcal{K}_l}\sum_{c\in\mathcal{C}_{{\rm N}_{l,k}}}\lambda_{l,k}^{(c)}=1 $ with the complementary slackness condition for all $ c\in\mathcal{C}_{{\rm N}_{l,k}} $
	\begin{equation}\label{the complementary slackness condition to t}
	\lambda_{l,k}^{(c)}\left[ t-p^{(c)}_{l,k}\mathsf{C}^{+}\left(\dfrac{1}{K^{(c)}_{{\rm N}_{l,k}}}+\mathsf{E}\left[\min_{i\in \mathcal{K}^{(c)}_{{\rm N}_{l,k}}}|h^{i\to k}_{l-1}|^2\right]P\right)\right]=0.
	\end{equation}
	
	Also, by setting the first derivative of $ \mathcal{L} $ with respect to $  p^{(c)}_{l,k} $ for all $ c\in\mathcal{C}_{{\rm N}_{l,k}} $, we have
	\begin{equation}\label{L respect to p}
	\lambda_{l,k}^{(c)}\mathsf{C}^{+}\left(\dfrac{1}{K^{(c)}_{{\rm N}_{l,k}}}+\mathsf{E}\left[\min_{i\in \mathcal{K}^{(c)}_{{\rm N}_{l,k}}}|h^{i\to k}_{l-1}|^2\right]P\right)-\mu=0
	\end{equation}
	with the complementary slackness condition $ \mu(\sum_{l=2}^{L}\sum_{k\in\mathcal{K}_l}\sum_{c\in\mathcal{C}_{{\rm N}_{l,k}}}p^{(c)}_{l,k}-1)=0 $.
	
	From Eq.~\eqref{L respect to p}, one can observe that $ \lambda_{l,k}^{(c)}=0 $,$ \forall  c\in\mathcal{C}_{{\rm N}_{l,k}}$ if $ \mu=0 $, which is contrary to $ \sum_{c\in\mathcal{C}_{{\rm N}_{l,k}}}\lambda_{l,k}^{(c)}=1 $. Thus, to obtain the optimal solution, $ \mu\neq0 $ should hold. For each $ c $ in $ \mathcal{C}_{{\rm N}_{l,k}} $, $p^{(c)}_{l,k} \mathsf{C}^{+}\left(\frac{1}{K^{(c)}_{{\rm N}_{l,k}}}+\mathsf{E}\left[\min_{i\in \mathcal{K}^{(c)}_{{\rm N}_{l,k}}}|h^{i\to k}_{l-1}|^2\right]P\right) $ should be the same and equal to $ t $ due to $ \mu\neq0 $, $ \lambda_{l,k}^{(c)}\neq0 $ and Eq.~\eqref{the complementary slackness condition to t}. Using $ \sum_{l=2}^{L}\sum_{k\in\mathcal{K}_l}\sum_{c\in\mathcal{C}_{{\rm N}_{l,k}}}\lambda_{l,k}^{(c)}=1 $ and 
	\begin{equation}\label{key}
	p^{(c)}_{l,k}=\dfrac{t}{{\mathsf{C}^{+}\left(\frac{1}{K^{(c)}_{{\rm N}_{l,k}}}+\mathsf{E}\left[\min_{i\in \mathcal{K}^{(c)}_{{\rm N}_{l,k}}}|h^{i\to k}_{l-1}|^2\right]P\right)}},
	\end{equation}
	the optimal $ t^{*} $ is given as
	\begin{equation}\label{Sad Op}
	\begin{split}
	t^{*}=&\left[\sum_{l=2}^{L}\sum_{k\in\mathcal{K}_{l}}\sum_{c\in\mathcal{C}_{{\rm N}_{l,k}}}\left[ \mathsf{C}^{+}\left(\dfrac{1}{K^{(c)}_{{\rm N}_{l,k}}}+\mathsf{E}\left[\min_{i\in \mathcal{K}^{(c)}_{{\rm N}_{l,k}}}|h^{i\to k}_{l-1}|^2\right]P\right)\right]^{-1}  \right]^{-1}
	\end{split}
	\end{equation}
	and the optimal $ {{p^{*}}^{(c)}_{l,k}} $ is given as
	\begin{equation}\label{key}
	{{p^{*}}^{(c)}_{l,k}}=\frac{ t^{*}}{\mathsf{C}^{+}\left(\frac{1}{K^{(c)}_{{\rm N}_{l,k}}}+\mathsf{E}\left[\min_{i\in \mathcal{K}^{(c)}_{{\rm N}_{l,k}}}|h^{i\to k}_{l-1}|^2\right]P\right)}.
	\end{equation}
	
	As a result, the computation rate with optimal time allocation and fixed power control is given as $ t^{*} $ (Eq.~\eqref{Sad Op}).	
	
	\begin{remark}[Special Cases]
		By setting $ L=2 $, $ K_2=1  $ and $ C_{{\rm N}_{2,1}}=1  $ in Eq.~\eqref{Sad Op}, it reduces to a simple case where $ K_1 $ nodes wish to compute a desired function at the fusion center directly as classical CoMAC mentioned in Section \ref{Aggregation Schemes}, and the rate of it, named the rate of CoMAC with fixed power control, is the same as Eq.~\eqref{RF-CoMAC with FPC}  \cite{jeon2014computation}. Also, by setting $ L=2 $, $ K_2=1  $, $ C_{{\rm N}_{2,1}}=K_1 $, and $ K^{(c)}_{{\rm N}_{2,1}}=1 $ in Eq.~\eqref{Sad Op}, it reduces to the time-sharing case as Eq.~\eqref{the time-sharing technique}.
	\end{remark}
	
	\subsection{Optimal Time Allocation and Adaptive Power Control}\label{Average Power Control s}
	We observe that each node in the hierarchical network is active only in the corresponding channel uses. To compute the functions more efficiently, long-term power control should be considered as $ \mathsf{E}\left[P^{i\to k}_{l-1}[m] \right]=P  $. The transmitted power of each node is set to
	\begin{equation}\label{Average Power Control}
	P^{i\to k}_{l-1}[m]=\left\lbrace 
	\begin{split}
	&c\dfrac{\min_{j\in \mathcal{K}^{(c)}_{{\rm N}_{l,k}}}|h^{j\to k}_{l-1}[m]|^2}{|h^{i\to k}_{l-1}[m]|^2}&,m\in\mathcal{T}_{l,k}^{(c)}\\
	&0&,{\rm otherwise}
	\end{split}\right..
	\end{equation}
	
	To satisfy the long-term power control constrain, we have
	\begin{equation}\label{key}
	\begin{split}
	\mathsf{E}\left[P^{i\to k}_{l-1}[m]\right]=&\sum\limits_{t=1}^{n}\Pr(m=t)P^{i\to k}_{l-1}[m]|_{m=t }\\
	\stackrel{(a)}{=}&\frac{c}{n}\sum_{t\in\mathcal{T}^{(c)}_{l,k}}\dfrac{\min_{j\in \mathcal{K}^{(c)}_{{\rm N}_{l,k}}}|h^{j\to k}_{l-1}[t]|^2}{|h^{i\to k}_{l-1}[t]|^2}\\
	\stackrel{(b)}{=}&c\alpha_{l,k}\beta_{l,k}^{(c)}\mathsf{E}\left[\dfrac{\min_{j\in \mathcal{K}^{(c)}_{{\rm N}_{l,k}}}|h^{j\to k}_{l-1}|^2}{|h^{i\to k}_{l-1}|^2} \right] 
	\end{split},
	\end{equation}
	which should be equal to $ P $. Then, $ c $ is obtained as
	\begin{equation}\label{value of c}
	c=\dfrac{P}{\alpha_{l,k}\beta_{l,k}^{(c)}\mathsf{E}\left[\dfrac{\min_{i\in \mathcal{K}^{(c)}_{{\rm N}_{l,k}}}|h^{i\to k}_{l-1}|^2}{|h^{j\to k}_{l-1}|^2} \right]}.
	\end{equation}
	
	Substituting Eqs.~\eqref{Average Power Control} and \eqref{value of c} into the rate in Theorem \ref{General Rate of Multi-Layer Network}, the computation rate is expressed as
	\begin{equation}\label{Grate of ML-FC with APC and OTA}
	\begin{split}
	R=&\min_{l\in[2:L]}\min_{k\in\mathcal{K}_{l}}\alpha_{l,k}\min_{c\in\mathcal{C}_{{\rm N}_{l,k}}} \beta^{(c)}_{l,k} \mathsf{C}^{+}\left(\dfrac{1}{K^{(c)}_{{\rm N}_{l,k}}}+\vphantom{\left.\frac{\mathsf{E}\left[ \min_{i\in\mathcal{K}^{(c)}_{{\rm N}_{l,k}}}|h^{i\to k}_{l-1}|^2\right] P}{\alpha_{l,k}\beta^{(c)}_{l,k}\mathsf{E}\left[{\min_{i\in\mathcal{K}^{(c)}_{{\rm N}_{l,k}}}|h^{i\to k}_{l-1}|^2}/{\left| h\right|^2 }\right]}\right) }\frac{\mathsf{E}\left[ \min_{i\in\mathcal{K}^{(c)}_{{\rm N}_{l,k}}}|h^{i\to k}_{l-1}|^2\right] P}{\alpha_{l,k}\beta^{(c)}_{l,k}\mathsf{E}\left[{\min_{i\in\mathcal{K}^{(c)}_{{\rm N}_{l,k}}}|h^{i\to k}_{l-1}|^2}/{\left| h\right|^2 }\right]}\right),
	\end{split}
	\end{equation}
	where $ h $ is used as a representative coefficient without loss of generality. 
	
	Considering adaptive power control, we formulate an optimization problem as Problem \ref{Average Power Control optimization} to maximize the computation rate.
	
	\begin{problem}\label{Average Power Control optimization}
		\begin{align}
		\mathop{\mathrm{maximize}}\limits_{\alpha_{l,k},\beta^{(c)}_{l,k}} & \quad \min_{l\in[2:L]}\min_{k\in\mathcal{K}_{l}}\alpha_{l,k}\min_{c\in\mathcal{C}_{{\rm N}_{l,k}}} \beta^{(c)}_{l,k} \mathsf{C}^{+}\left(\dfrac{1}{K^{(c)}_{{\rm N}_{l,k}}}+\vphantom{\quad\left.\frac{\mathsf{E}\left[ \min_{i\in\mathcal{K}^{(c)}_{{\rm N}_{l,k}}}|h^{i\to k}_{l-1}|^2\right] P}{\alpha_{l,k}\beta^{(c)}_{l,k}\mathsf{E}\left[{\min_{i\in\mathcal{K}^{(c)}_{{\rm N}_{l,k}}}|h^{i\to k}_{l-1}|^2}/{\left| h\right|^2 }\right]}\right) \nonumber} \quad\frac{\mathsf{E}\left[ \min_{i\in\mathcal{K}^{(c)}_{{\rm N}_{l,k}}}|h^{i\to k}_{l-1}|^2\right] P}{\alpha_{l,k}\beta^{(c)}_{l,k}\mathsf{E}\left[{\min_{i\in\mathcal{K}^{(c)}_{{\rm N}_{l,k}}}|h^{i\to k}_{l-1}|^2}/{\left| h\right|^2 }\right]}\right) \nonumber \\
		{\rm s.t.}&\quad  \sum_{l=2}^{L}\sum_{k\in\mathcal{K}_l}\alpha_{l,k}=1\\
		&\quad  \sum_{c\in\mathcal{C}_{{\rm N}_{l,k}}}\beta^{(c)}_{l,k}=1,\forall l\in[2:L], \forall k\in\mathcal{K}_l
		\end{align}
	\end{problem}
	
	By introducing the convex relaxation as  $ p_{l,k}^{(c)} = \alpha_{l,k}\beta_{l,k}^{(c)}$, this problem is rewritten as the following form.
	\begin{problem}\label{rewritten Average Power Control optimization}		
		\begin{align}
		&\mathop{\mathrm{maximize}}\limits_{p^{(c)}_{l,k},t}  \quad t  \nonumber \\
		&{\rm s.t.}\ p^{(c)}_{l,k}\mathsf{C}^{+}\left(\dfrac{1}{K^{(c)}_{{\rm N}_{l,k}}}+\frac{\mathsf{E}\left[\min_{i\in\mathcal{K}^{(c)}_{{\rm N}_{l,k}}}|h^{i\to k}_{l-1}|^2\right]P}{p^{(c)}_{l,k}\mathsf{E}\left[{\min_{i\in\mathcal{K}^{(c)}_{{\rm N}_{l,k}}}|h^{i\to k}_{l-1}|^2}/{\left| h\right|^2 }\right]}\right)\ge t,\nonumber\\
		&\quad\quad\quad\quad\quad\quad\quad\forall l\in[2:L], \forall k \in\mathcal{K}_{l}, \forall c \in \mathcal{C}_{l,k} \label{convex constrains}\\
		&\quad\quad  \sum_{l=2}^{L}\sum_{k\in\mathcal{K}_l}\sum_{c\in\mathcal{C}_{{\rm N}_{l,k}}}p_{l,k}^{(c)}=1 \nonumber
		\end{align}
	\end{problem}
	
	Problem \ref{rewritten Average Power Control optimization} now is a convex problem since the constrain, Eq.~\eqref{convex constrains}, is concave. Hence, the above optimization problem has a unique maximum. The Lagrangian function is given as
	\begin{equation}\label{Lagrangian}
	\begin{split}
	\mathcal{L}=&t-\sum_{l=2}^L\sum_{k\in\mathcal{K}_{l}}\sum_{c\in\mathcal{C}_{{\rm N}_{l,k}}}\lambda_{l,k}^{(c)}\left[t-p^{(c)}_{l,k}\mathsf{C}^{+}\left(\dfrac{1}{K^{(c)}_{{\rm N}_{l,k}}}+\vphantom{\left.\left.\frac{\mathsf{E}\left[\min_{i\in\mathcal{K}^{(c)}_{{\rm N}_{l,k}}}|h^{i\to k}_{l-1}|^2\right]P}{p^{(c)}_{l,k}\mathsf{E}\left[{\min_{i\in\mathcal{K}^{(c)}_{{\rm N}_{l,k}}}|h^{i\to k}_{l-1}|^2}/{\left| h\right|^2 }\right]}\right) \right]}\frac{\mathsf{E}\left[\min_{i\in\mathcal{K}^{(c)}_{{\rm N}_{l,k}}}|h^{i\to k}_{l-1}|^2\right]P}{p^{(c)}_{l,k}\mathsf{E}\left[{\min_{i\in\mathcal{K}^{(c)}_{{\rm N}_{l,k}}}|h^{i\to k}_{l-1}|^2}/{\left| h\right|^2 }\right]}\right) \right]\\
	&-\mu\left(\sum_{l=2}^{L}\sum_{k\in\mathcal{K}_l}\sum_{c\in\mathcal{C}_{{\rm N}_{l,k}}}p_{l,k}^{(c)}-1\right)
	\end{split}
	\end{equation}
	with the complementary  slackness condition
	\begin{equation}\label{key}
	\mu\left(\sum_{l=2}^{L}\sum_{k\in\mathcal{K}_l}\sum_{c\in\mathcal{C}_{{\rm N}_{l,k}}}p_{l,k}^{(c)}-1\right)=0,
	\end{equation}
	where $ \left\lbrace \lambda_{l,k}^{(c)}\right\rbrace  $ and $ \mu $ are Lagrange multipliers.
	
	We apply the KKT optimality conditions to the Lagrangian function to obtain the optimal factor $ {p^{*}}^{(c)}_{l,k} $. By setting the first derivative of $ \mathcal{L} $ as Eq.~\eqref{Lagrangian} with respect to $ {p}^{(c)}_{l,k} $ to zero, we have
	\begin{equation}\label{key}
	\ln\left(\dfrac{1}{K^{(c)}_{{\rm N}_{l,k}}}+\dfrac{\varepsilon^{(c)}_{l,k}}{{p}^{(c)}_{l,k}}\right)-\dfrac{\varepsilon^{(c)}_{l,k}}{{p}^{(c)}_{l,k}\left( \dfrac{1}{K^{(c)}_{{\rm N}_{l,k}}}+\dfrac{\varepsilon^{(c)}_{l,k}}{{p}^{(c)}_{l,k}}\right) }=\dfrac{\mu\ln(2)}{\lambda_{l,k}^{(c)}},
	\end{equation}
	where $ \varepsilon^{(c)}_{l,k}=\frac{\mathsf{E}\left[\min_{i\in\mathcal{K}^{(c)}_{{\rm N}_{l,k}}}|h^{i\to k}_{l-1}|^2\right]P}{\mathsf{E}\left[{\min_{i\in\mathcal{K}^{(c)}_{{\rm N}_{l,k}}}|h^{i\to k}_{l-1}|^2}/{\left| h\right|^2 }\right]} $.
	
	Then, each optimal factor is expressed as
	\begin{equation}\label{OPeta}
	{p^{*}}^{(c)}_{l,k}=\max\left\lbrace 0,-\varepsilon^{(c)}_{l,k}K^{(c)}_{{\rm N}_{l,k}}\left[ 1+\left( {\tau}^{(c)}_{l,k}\right)^{-1}  \right]^{-1} \right\rbrace,
	\end{equation}
	where $  {\tau}^{(c)}_{l,k} $ is a Lambert $ W $ function as
	\begin{equation}\label{Vg}
	{\tau}^{(c)}_{l,k}=W\left(-{2^{-\frac{\mu}{\lambda_{l,k}^{(c)}}}}{\left( K^{(c)}_{{\rm N}_{l,k}}\right)^{-1}}\exp(-1)\right),
	\end{equation}
	while  $ {p^{*}}^{(c)}_{l,k} $ satisfies 
	\begin{equation}\label{Constrain}
	\left\lbrace
	\begin{split}
	\sum_{l=2}^{L}\sum_{k\in\mathcal{K}_l}\sum_{c\in\mathcal{C}_{{\rm N}_{l,k}}}{p^{*}}^{(c)}_{l,k}\le1,\quad\mu=0\\
	\sum_{l=2}^{L}\sum_{k\in\mathcal{K}_l}\sum_{c\in\mathcal{C}_{{\rm N}_{l,k}}}{p^{*}}^{(c)}_{l,k}=1,\quad\mu>0
	\end{split}
	\right..
	\end{equation} 
	
	\begin{remark}[Special Cases]
		By setting $ L=2 $, $ K_2=1  $ and $ C_{{\rm N}_{2,1}}=1  $ in Eq.~\eqref{Grate of ML-FC with APC and OTA}, the rate of it is the same as the rate
		\begin{equation}\label{RF-CoMAC with APC}
		R=\mathsf{C}^+\left( \dfrac{1}{K_1}+\dfrac{\mathsf{E}\left[\min_{i \in [1:K_1]} |h_{1,i}|^2\right]P }{\mathsf{E}\left[{\min_{i \in [1:K_1]}|h_{1,i}|^2}/{|h|^2} \right]}\right)
		\end{equation}
		in  \cite[Theorem 5]{jeon2014computation} as the rate of CoMAC with adaptive power control. Also, by setting $ L=2 $, $ K_2=1  $, $ C_{{\rm N}_{2,1}}=K_1 $, $ K^{(c)}_{{\rm N}_{2,1}}=1 $ and $ \beta^{(c)}_{l,k}=\frac{1}{K_1} $ in Eq.~\eqref{Grate of ML-FC with APC and OTA} as the time-sharing case, an improved rate is obtained as $ R=\frac{1}{K_1}\mathsf{E}\left[ \mathsf{C}\left(\left|h \right|^2K_1P  \right) \right] $ compared with Eq.~\eqref{the time-sharing technique}.
	\end{remark}

	\section{Simulation Results and Discussion}\label{Simulation Results and Discussions}
	
	In this section, we provide simulation results of the computation rates of ML-FC, the time-sharing scheme as Eq.~\eqref{the time-sharing technique}, CoMAC with fixed power control as Eq.~\eqref{RF-CoMAC with FPC} and CoMAC with adaptive power control as Eq.~\eqref{RF-CoMAC with APC}. In our simulation, the average signal-to-noise ratio (SNR) is the same as $ P $ because the variance of the noise is set as one. We consider i.i.d. Rayleigh fading channel, i.e., the exponential distribution with parameter one. The abbreviations for fixed power control, adaptive power control, average time allocation, and optimal time allocation are FPC, APC, ATA, and OTA, respectively.
	
	\begin{figure}
		\centering
		\includegraphics[width=\linewidth]{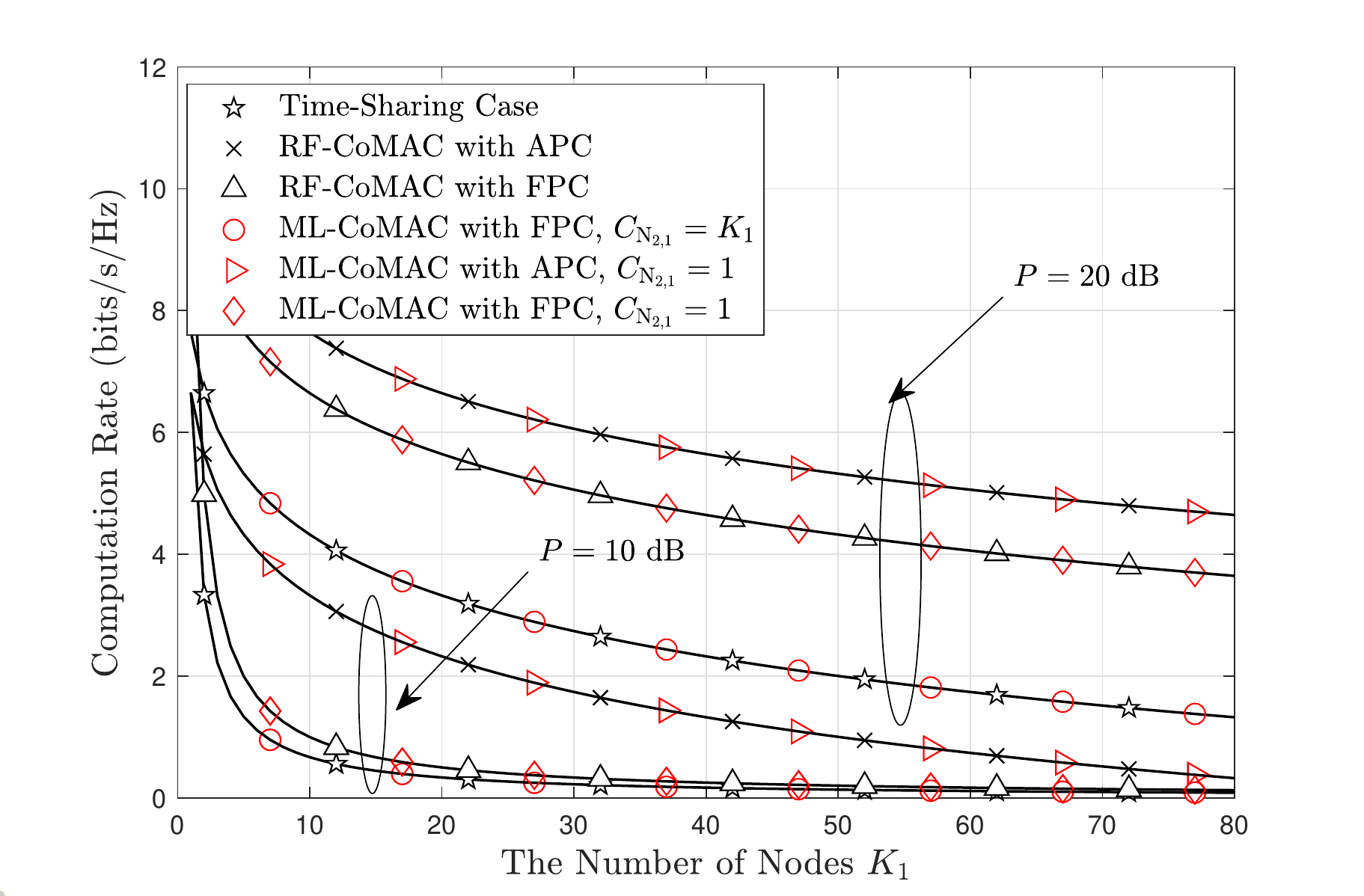}
		\caption{Computation rates of CoMAC with different schemes with respect to the number of source nodes $ K_1 $ and $ P $ when $ L=2 $.}
		\label{fig:cs1}
	\end{figure}
	
	\begin{figure}
		\centering
		\includegraphics[width=\linewidth]{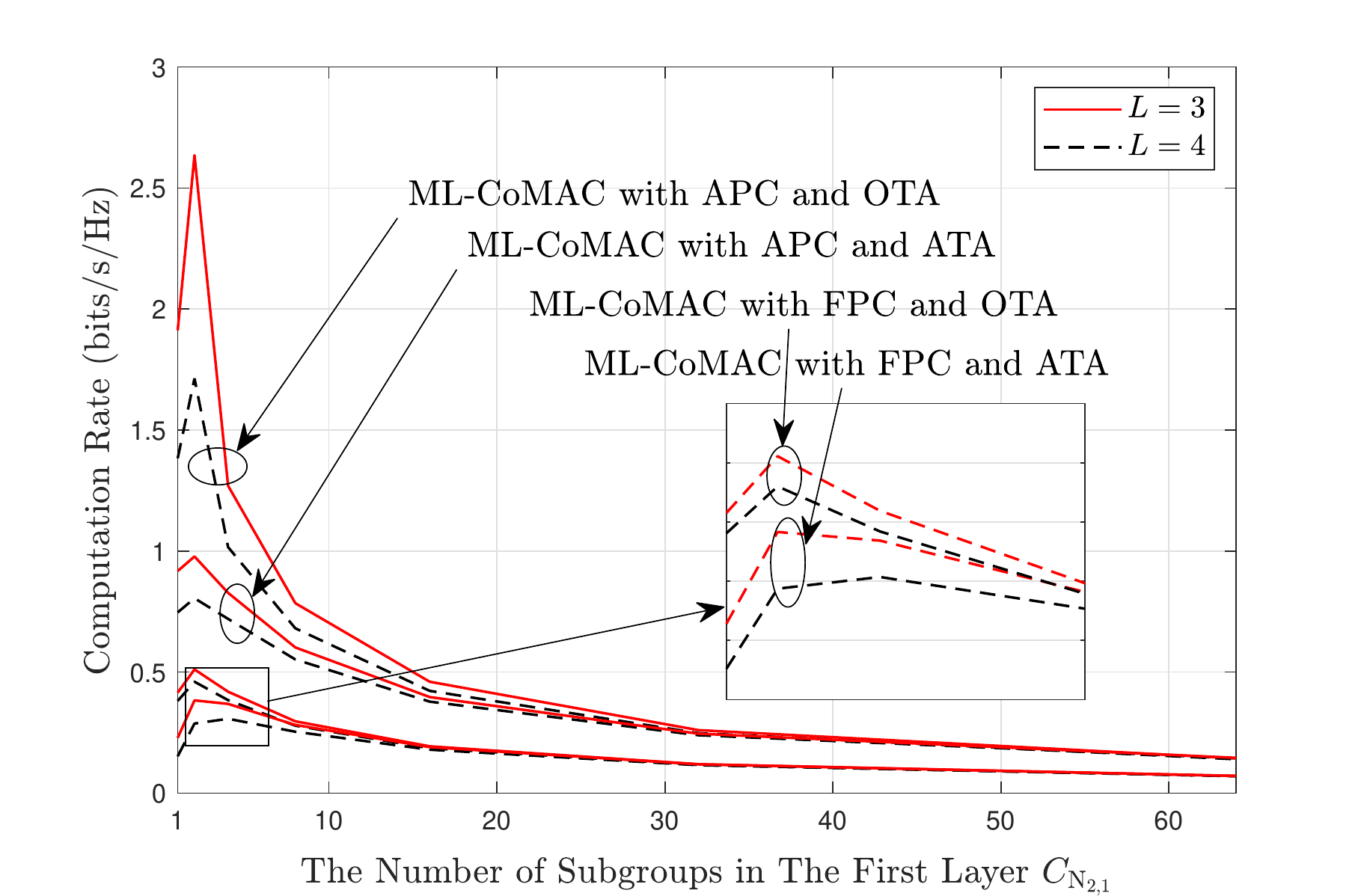}
		\caption{Computation rates of ML-FC with different schemes with respect to the number of subgroups $ C_{{\rm N}_{2,1}} $ and the number of layers $ L $ when $ K_1 $=64.}
		\label{fig:cs2}
	\end{figure}
	
	\begin{figure}
		\centering
		\includegraphics[width=\linewidth]{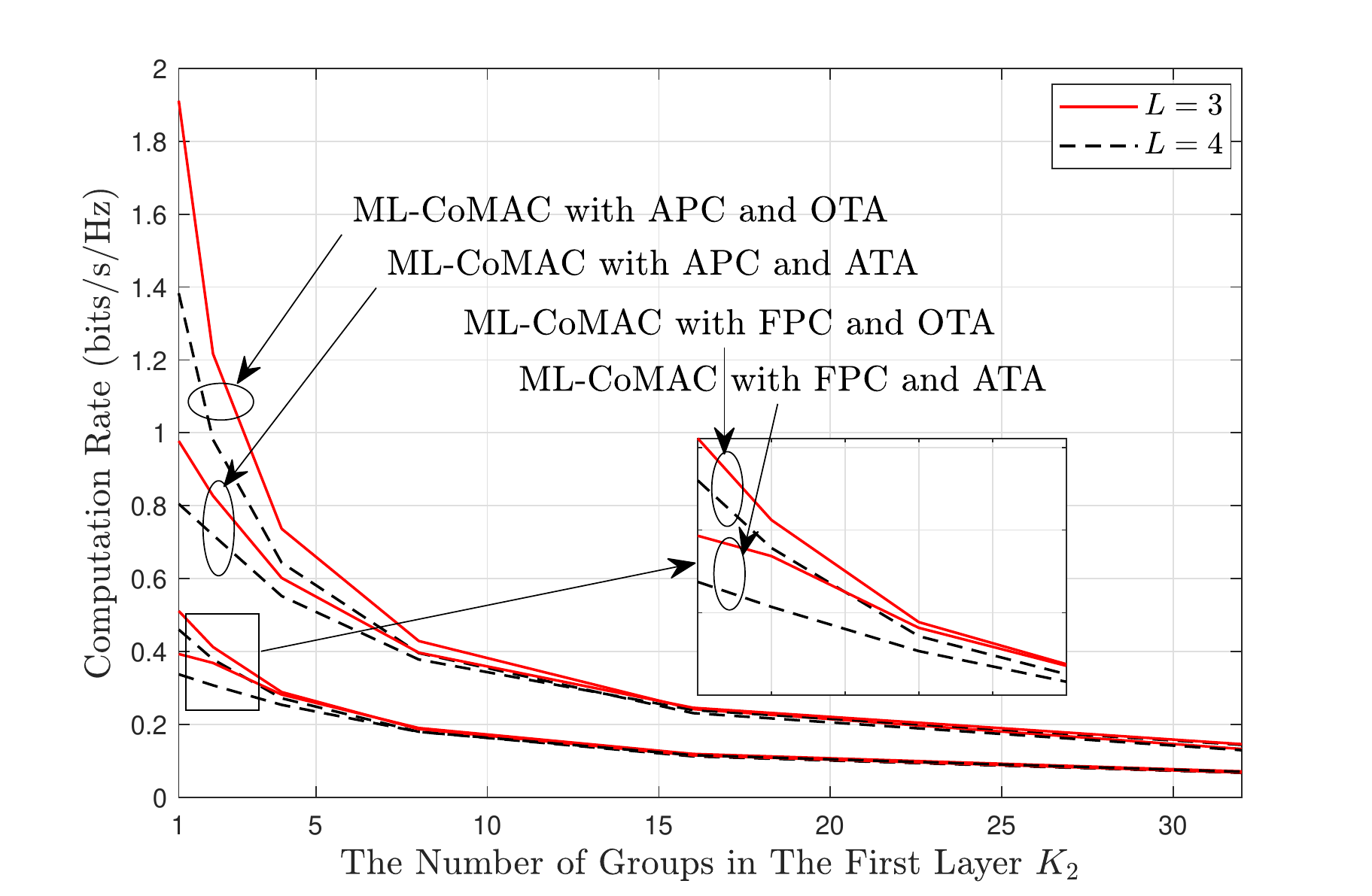}
		\caption{Computation rates of ML-FC with different schemes with respect to the number of subgroups $ K_2 $ and the number of layers $ L $ when $ K_1=64 $.}
		\label{fig:cs3}
	\end{figure}
	
	\begin{figure}
		\centering
		\includegraphics[width=\linewidth]{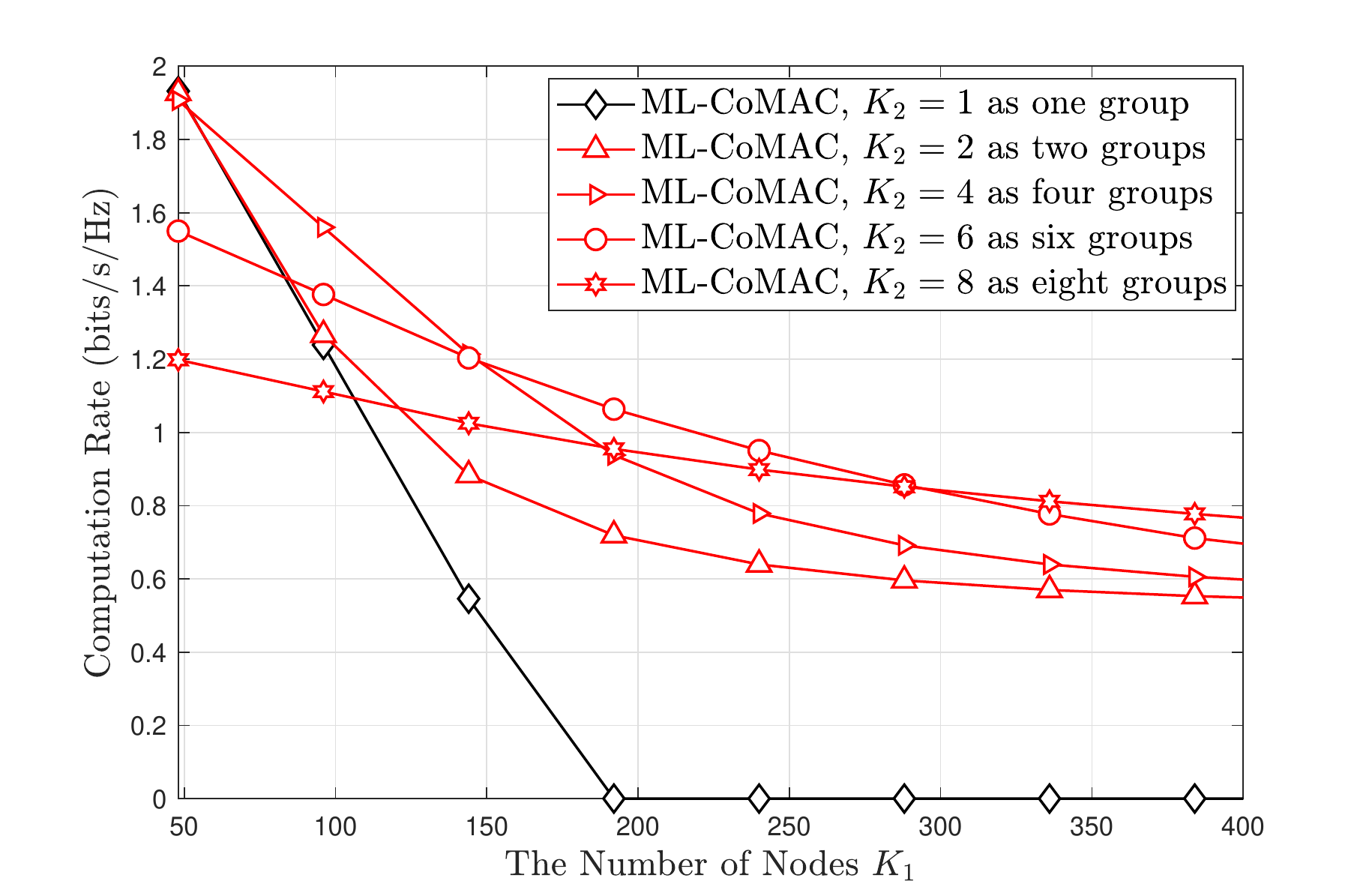}
		\caption{Computation rates of ML-FC with respect to the number of groups $K_2 $ and the number of source nodes $ K_1 $ when $ L=3 $.}
		\label{fig:cs4}
	\end{figure}
	
	Since the hierarchical network is a more general case compared with the relay-free network, the rates of CoMAC (Eqs.~\eqref{the time-sharing technique}, \eqref{RF-CoMAC with FPC} and \eqref{RF-CoMAC with APC}) should be generalized by the rates of ML-FC with different parameters. Thus, in Fig.~\ref{fig:cs1}, their relationship is given. By setting $ L=2 $, the hierarchical network reduces to the relay-free network aiming at computing the desired function associated with $ K_1 $ source nodes directly. When the number of subgroups $ C_{{\rm N}_{2,1}} $ is equal to the number of source nodes $ K_1 $, the fusion center collects all the individual data from $ K_1 $ nodes as the time-sharing case and the rate of ML-FC with FPC is the same as Eq.~\eqref{the time-sharing technique} by setting $ C_{{\rm N}_{2,1}}=K_1 $ in Eq.~\eqref{Sad Op}. By setting $ C_{{\rm N}_{2,1}}=1 $ in Eq.~\eqref{Sad Op}, all the nodes transmit signals simultaneously to the fusion center as the number of the subgroups is 1. It generalizes the rate of CoMAC with FPC ( Eq.~\eqref{RF-CoMAC with FPC}). Similarly, by setting $ C_{{\rm N}_{2,1}}=1 $ in Eq.~\eqref{Grate of ML-FC with APC and OTA}, the rate of CoMAC with APC (Eq.~\eqref{RF-CoMAC with APC}) is obtained.
	
	The computation rates of ML-FC with different schemes versus the number of subgroups in the first layer $ C_{{\rm N}_{2,1}} $ and the number of layers $ L $ are demonstrated in Fig.~\ref{fig:cs2}. In this case, we consider the hierarchical network where $ K_1=64 $ nodes are deployed in the first layer and each of the rest layers owns one node. One can observe that the computation rate decreases as the number of layers increases. Since each layer has to be allocated some channel uses to compute the corresponding functions, i.e., subgroup functions and group functions, the increase in the number of layers causes the decrease in the number of channel uses allocated to each layer when the total of channel uses is fixed. Besides, the computation rate is improved by setting $ C_{{\rm N}_{2,1}} =2 $. This implies that the group function should be divided into several subgroup functions to be computed instead of computing it directly. Compared with ML-FC with FPC, ML-FC with APC improves the rate. Also, optimal time allocation provides further improvement.

	However, the impact of the number of groups is different from the impact of the number of subgroups. In Fig.~\ref{fig:cs3}, we show the computation rates of ML-FC for different schemes versus the number of groups in the first layer $ K_{2} $\footnote{As demonstrated in Section \ref{Hierarchical Networks}, the sum of the number of groups in the $ l $-th layer is equal to the number of nodes in the $ (l+1) $-th layer since each node in the $ (l+1) $-th is allocated a group from the $ l $-th layer in the hierarchical network.}. The main difference from Fig.~\ref{fig:cs2} is that the increase in the groups results in the worse performance since each group is allocated fewer channel uses when the channel uses and $ C_{{\rm N}_{2,1}} $ are fixed. With fewer channel uses, the number of the group functions computed at the corresponding node is fewer. Thus, the computation rate of ML-FC decreases.

	Although Fig.~\ref{fig:cs3} suggests that the number of groups in a network should be as few as possible, it does not mean that the increase in the number of groups only has disadvantage. As shown in Fig.~\ref{fig:cs4}, we simulate the computation rates with respect to the number of groups in the first layer and the number of the source nodes in the first layer. One can observe that all the rates decrease as the number of source nodes $ K_1 $ increases. Also, when $ K_1 $ is small, the relation between the rate and the number of groups is the same as that in Fig.~\ref{fig:cs3}. However, as $ K_1 $ becomes larger, unlike the rate of ML-FC with one group decreasing rapidly, the rates of ML-FC with multiple groups keep a slower decrease. Especially, ML-FC with eight groups provides the slowest decrease, which implies that ML-FC with more groups can support a network with more nodes. Thus, it provides a way to design a network that can afford massive numbers of nodes by increasing the number of groups in this network.

	\section{Conclusion}
	\label{Conclusion}
	In this paper, we have combined the uses of CoMAC and orthogonal communication to attain the computation of functions in the disorganized network. First, we have reorganized the disorganized network into the hierarchical network including multiple layers, which consists of subgroups and groups. In the hierarchical network, ML-FC has been developed where subgroup functions and group functions are obained by CoMAC and orthogonal communication, respectively. Then, the desired function at the fusion center is reconstructed by these subgroup and group functions. To reliably reconstruct the desired function over multiple layers, we have characterized the relationship among subgroup functions, group functions, and desired functions. With the given relationship, we have derived the general computation rate of ML-FC, which suggests that the computation rate is determined by the subgroup function with the worst rate. Furthermore, we have formulated optimization problems taking into account time allocation and power control. The closed-form optimal solutions have been given with respect to different cases, which generalizes the existing CoMAC works.

		\bibliography{NOAH}
	\bibliographystyle{IEEEtran}
\end{document}